\documentclass[%
 reprint,
 amsmath,amssymb,
 aps,longbibliography,superscriptaddress
]{revtex4-1}

\usepackage{graphicx}
\usepackage{dcolumn}
\usepackage{subfigure}
\usepackage{bm}

\begin{document}


\title{Criticality and Phase Diagram of 
Quantum Long-Range $\text{O(N)}$ models}

\author{Nicol\`o Defenu}
\affiliation{Institut f\"ur Theoretische Physik, Universit\"at 
Heidelberg, D-69120 Heidelberg, Germany}

\author{Andrea Trombettoni}
\affiliation{CNR-IOM DEMOCRITOS Simulation Center, Via Bonomea 265, 
I-34136 Trieste, Italy}
\affiliation{SISSA, Via Bonomea 265, I-34136 Trieste, Italy}
\affiliation{INFN, Sezione di Trieste, I-34151 Trieste, Italy}

\author{Stefano Ruffo}
\affiliation{SISSA, Via Bonomea 265, I-34136 Trieste, Italy}
\affiliation{INFN, Sezione di Trieste, I-34151 Trieste, Italy}


\date{\today}

\begin{abstract}
Several recent experiments in atomic, molecular and optical systems 
motivated a huge interest in the study of quantum long-range 
systems. 
Our goal in this paper is to present a 
general description of their critical behavior and phases, 
devising a treatment valid 
in $d$ dimensions, with an exponent $d+\sigma$ for the power-law 
decay of the couplings in the presence of an $O(N)$ symmetry. 
By introducing a convenient ansatz for the effective action, 
we determine the phase diagram for the $N$-component quantum rotor model 
with long-range interactions, 
with $N=1$ corresponding to the Ising model. The phase diagram in 
the $\sigma-d$ plane shows a non trivial dependence on $\sigma$. 
As a consequence 
of the fact that the model is quantum, 
the correlation functions are anisotropic in the spatial 
and time coordinates for $\sigma$ smaller than a critical value and in this 
region the isotropy is not restored even at criticality. Results 
for the correlation length exponent $\nu$, the dynamical critical exponent 
$z$ and a comparison with numerical findings for them are presented.
\end{abstract}

\pacs{Valid PACS appear here}
\maketitle


\section{Introduction}
\label{sec:introduction}

A series of remarkable advancements in the last decade lead to 
a striking development of the experimental techniques for the control 
and manipulation of atomic, 
molecular and optical systems 
such as Rydberg atoms \cite{Saffman2010}, 
dipolar quantum gases \cite{Lahaye2009}, polar molecules \cite{Ritsch2013}, 
multimode cavities \cite{Carr2009} and trapped ions \cite{Blatt2012}. 
Such progress paved the way for the simulation 
of a variety of quantum models \cite{Bloch2008}, and in 
this direction one of the most active field of research is the 
implementation and study of equilibrium and dynamical properties of 
quantum long-range (LR) models 
\cite{Britton2012,Schauss2012,Aikawa2012,Lu2012,Yan2013,Islam2013,Richerme2014,Jurcevic2014,Douglas2015,Schempp2015,Landig2015,Landig2015a}. 
Ising and/or $XY$ quantum spin chains with tunable LR interactions 
can currently be realized using $Be$ ions stored in a Penning trap 
\cite{Britton2012}, neutral atoms coupled to
photonic modes of a cavity \cite{Douglas2015,Landig2015,Landig2015a} or 
with trapped ions coupled to 
motional degrees of freedom \cite{Islam2013,Richerme2014,Jurcevic2014}. 
A key property in these studies
is that the resulting interactions decay algebraically with 
the distance $r$ and that the decay exponent can be experimentally tuned 
\cite{Islam2013,Richerme2014,Jurcevic2014}. As an example, 
Rydberg gases have been 
used to observe and study spatially ordered structures 
\cite{Schauss2012} and correlated transport \cite{Schempp2015}.  
Dipolar spin-exchange 
interactions with lattice-confined polar molecules were observed 
as well \cite{Yan2013}. The possibility of control LR interactions 
in spin chains simulated with trapped ions was also at the basis 
of the recent experimental simulation of the $1D$ Schwinger model 
\cite{Martinez2016}.


An important reason for the interest in the equilibrium and non-equilibrium 
properties of quantum LR systems is the connection with typical themes 
of classical LR physics. The traditional interest in the behavior of 
LR interacting statistical mechanics models 
has been largely due by the outnumbering possible applications 
in condensed matter, plasma physics, astrophysics and cosmology 
\cite{Dauxois2010,Campa2014}. Therefore, the general question is
how quantum fluctuations modify the traditional picture of LR interactions
into complex systems. 
We focus on the study of the criticality 
in quantum LR systems and 
the development of a general renormalization group (RG) 
approach for their study.

From the point of view of critical behavior, the main
difference between classical and quantum systems is due to the
presence of unitary quantum dynamics in the latter case. The 
critical behavior in the time domain is usually described in terms
of the dynamical critical exponent $z$, which is related 
to the critical scaling $\Delta \propto L^{-z}$ of the gap $\Delta$ 
in the thermodynamic limit $L \to \infty$, where $L$ is the linear
size of the system. This is different from classical systems, 
where the value of $z$ depends on the chosen dynamics.

{In several short-range (SR) systems the value of this exponent is 
strictly one, leading to the well known equivalence between the 
universality class of a $d$ dimensional quantum system and its classical 
equivalent in $d_{\text{cl}}=d+1$ dimensions,
where $d_{\text{cl}}$ is the dimension of the corresponding classical system. 
For non-unitary values of the dynamical critical exponent $z$,
which is the case of LR interacting quantum systems, the relation between
classical and quantum critical behavior is more subtle. The general 
expectation is 
that classical and quantum universalities should be connected 
in general for $d_{\text{cl}}=d+z$ \cite{Sachdev2011}.}

{The aim of our investigation is focused on the derivation
of the critical exponent $z$ and of its spatial counterpart, i.e. the
correlation length critical exponent $\nu$. Various relations, exact or not, 
which connect both classical and quantum, SR and
LR systems, are also investigated. In order to accomplish 
this task we employ the functional renormalization group (FRG) 
formalism already developed for classical
isotropic and anisotropic LR $O(N)$ models \cite{Defenu2015,Defenu2016}.} 

The interplay between LR interactions and quantum
effects is a traditional topic in condensed matter physics. 
A major example is provided by the $\frac{1}{r^{2}}$
Ising model \cite{Thouless1969}, displaying 
a behavior related to the spin-$\frac{1}{2}$ Kondo problem 
with the occurrence of a 
topological phase transition of the Berezinskii-Kosterlitz-Thouless 
(BKT) type \cite{Yuval1970,Cardy1981}.
Moreover, the experimental realization of dipolar Ising spin glasses in the 
$90$'s led to investigations of the behavior of quantum LR 
Ising and rotor models \cite{Dutta2001}.

The application of functional 
RG techniques to classical LR spin systems 
 produced a nice all-in-one picture of the phase 
transitions occurring in these models, contributing
to clarify the behavior of the anomalous dimension
of these systems \cite{Defenu2015,Sak1973,Picco2012,Blanchard2013,Brezin2014}.
Moreover functional RG formalism was successfully applied to the case of
anisotropic LR interactions \cite{Defenu2016}. 
Good agreement with Monte Carlo (MC) simulations was found, 
not only in the two dimensional 
Ising model \cite{Luijten2002,Angelini2014,Mori2010,Horita2016}, but also
in $1$-dimensional LR bond percolation by means of effective dimension
relations \cite{Gori2016}. 

Comparing with traditional perturbative RG analysis, functional
RG was able to produce, at least for SR $O(N)$ models, numerical values
for the critical exponents, whose accuracy remains stable as a function of
all relevant system parameters, i. e. the dimension $d$ and number of components  $N$. In the classical
SR $O(N)$ models  the accuracy never falls below 20\%, while remaining well below 10\%
for all continuos symmetries $N\geq 2$ \cite{Codello2013,Codello2015}. Such accuracy is expected to hold also in the LR case, where correlation effects should never
be stronger than in the two dimensional classical SR Ising model. 

The issue we address is the general 
description of the criticality 
in quantum LR spin models and the development of a RG framework for this 
investigation. As a first, one would naively think that the problem 
is reduced to the study of a LR classical problem, but in this paper it is
shown explicitly that this is the case only for $\sigma$ large enough. 
A main aim of this paper is therefore 
to widen and unify the theoretical picture 
for the phase transitions quantum LR models 
and to construct the full landscape 
of their phase diagram within a single formalism. We also 
determine the dynamical critical exponent $z$, which 
characterizes the critical quantum dynamics. 

The structure of the paper 
is the following. In Section \ref{sec:2} we introduce the models 
used in the rest of the paper. In Section \ref{sec:3} we discuss the 
field theoretical description and the RG methods, and we present 
our ansatz for the effective action. We also derive the mean field results 
and study the corrections to the phase diagram due to the effects 
of the presence of anomalous dimension. Estimates for the critical 
exponents in $d=1$ and $d=2$ are presented in Section \ref{sec:4}, 
together with a discussion of the comparison with some numerical results 
available in literature. Our conclusions are presented in Section 
\ref{sec:5}, together with a summary of the relevance and implications 
of our results.

\section{Long-Range models}\label{sec:2}

In this paper we consider two models, namely 
the Ising model in a transverse field 
and the quantum rotor model. The Ising Hamiltonian 
in a transverse field ${\cal H}$ reads
\begin{align}
\label{Eq1}
H_{{\rm I}}=-\sum_{ij}\frac{J_{ij}}{2}\sigma_{i}^{z}\sigma_{j}^{z}-
{\cal H}\sum_{i}\sigma_{i}^{x},
\end{align}
where the $\sigma^{z}$ and $\sigma^{x}$ 
are the Pauli matrices defined in the sites $i,j$ of a $d$-dimensional 
hyper-cubic lattice. The quantum rotor model can be written as 
\begin{align}
\label{Eq2}
H_{{\rm R}}=-\sum_{ij}\frac{J_{ij}}{2}\hat{\boldsymbol{n}}_{i}\cdot\hat{\boldsymbol{n}}_{j}+\frac{\lambda}{2}\sum_{i}\mathcal{L}_{i}^{2}, 
\end{align}
where the $\hat{\boldsymbol{n}}_i$ are $N$ components unit length vector 
operators ($\hat{\boldsymbol{n}}_{i}^{2}=1$), $\lambda$ is a real constant 
and $\mathcal{L}$ is the invariant
operator formed from the asymmetric rotor space angular momentum tensor 
\cite{Sachdev2011}. 
The interaction matrix in both cases is power-law decaying with the distance
\begin{align}
\label{Eq3}
J_{ij}=\frac{J}{r_{ij}^{d+\sigma}}
\end{align}
with $J$ a positive constant, $d$ the spatial dimension of the system and $\sigma$ any real number, while $r_{ij}$ is the distance between the sites $i$ and $j$. 

The interaction potential in Eq.\eqref{Eq3} 
induces in the model radically different
behaviors depending on the value of the exponent $\sigma$. 
In particular, for $\sigma<0$ the internal energy of 
the system may diverge
in the thermodynamic limit, leading to an ill defined model. 
However, a suitable redefinition of the 
interaction strength \cite{Kac1963} produces finite interaction energies, 
still preserving many interesting results typical of non additive
systems \cite{Campa2014}. 

For fast enough decaying interactions $\sigma > 0$ 
the system is always additive and  thermodynamics is well defined. 
For $\sigma$ lying into a certain range, the finite temperature system 
may present
a phase transition with spontaneous symmetry breaking at
a finite critical temperature $T_{c}$. 

For a classical spin system the phase diagram in the $(d,\sigma)$ plane
is therefore divided into three regions \cite{Sak1973}.  
For $2\sigma \le d$ 
the universal behavior is the one obtained at mean field level, 
while for $\sigma$ larger than a critical 
value $\sigma^{*}$ the system has the same critical behavior of its SR analogue.
Finally for $d/2 < \sigma \le \sigma^{*}$ LR interactions are relevant 
and the system
has a phase transition with peculiar LR critical behavior. 

The determination of the boundary $\sigma^{*}$ 
where LR interactions become irrelevant respect to the usual SR 
ones has been at the center of intense investigations in last few years
\cite{Brezin2014,Angelini2014,Defenu2015}, see as well the recent works 
\cite{Behan2017,Behan2017a}. The functional RG framework 
\cite{Tetradis1992,Morris1994,Berges2002} proved to be useful 
to explore the effect of LR interactions \cite{Defenu2015,Defenu2016} 
in a compact way.
The accuracy of such results remains reliable 
for all values of the system parameters, i.e. 
the spatial dimension $d$, the number of components $N$ and the 
criticality index $i$.

Rotor models are the straightforward generalization 
to the quantum case of the classical $O(N)$ models \cite{Sachdev2011}. 
Their low energy behavior describes the physics of many relevant physical 
models, in particular the $N=3$ case is in the same universality of 
antiferromagnetic quantum Heisenberg spin systems, while 
the $N=2$ case is related to the 
quantum critical behavior of the Bose-Hubbard model. 

In this paper we only consider the case $\sigma>0$ in order to
have a well defined thermodynamics \cite{Dauxois2010,Campa2014}. The system
undergoes a paramagnetic to ferromagnetic quantum phase transition (QPT)
at zero temperatures for all $d$ and $\sigma$ in the Ising case, 
equation \eqref{Eq1},
i.e. the one component $N=1$ case of the rotor model in equation \eqref{Eq2}.
On the other hand, for $N\geq 2$, due to the Mermin-Wagner theorem (MWT)
the system manifests spontaneous symmetry breaking (SSB) only for
$\sigma<\sigma_{*}$ in $d=1$, where $\sigma_{*}$ is the threshold value above 
which short-range (SR) behavior is recovered.

At finite temperature $T>0$ the system has a finite temperature 
classical phase transition (CPT), which lies in the 
same universality class of standard
LR classical $O(N)$ spin systems. The results for the finite 
temperature transition
behavior were subject of intense, long-lasting investigations 
in the last decades 
\cite{Sak1973,Defenu2015} and they will be used, but not further discussed, 
in this paper. 

\section{Field theoretical representation and phase diagram}\label{sec:3}

The universal behavior of condensed matter and statistical mechanics 
models can be investigated by means of field
theoretical techniques \cite{Mussardo2010}. 
In order to employ the functional RG approach,
one may consider the Wetterich equation \cite{Wetterich1993}
\begin{align}
\label{Eq4}
\partial_{t}\Gamma_{k}=\frac{1}{2}{\rm Tr}\left[\frac{\partial_{t}R_{k}}{\Gamma^{(2)}+R_{k}}\right],
\end{align} 
where $R_{k}$ is an infrared regulator function, 
$k$ a finite scale proportional to the
inverse system size, $k\propto L^{-1}$, and 
$t=\log\left(\frac{k}{k_{0}}\right)$ the logarithmic scale. 
$\Gamma$ is the effective action of the system and plays the role of an 
exact Ginzburg-Landau free
energy, while $\Gamma^{(2)}$ 
is its second derivative with respect to the system order parameters.

In order to solve equation \eqref{Eq4} it is useful 
to restrict ourselves to a functional space spanned
by a finite number of functions and/or couplings. Using 
the field theory representation and the Trotter decomposition, 
it can be shown that the models defined in equations \eqref{Eq1} 
and \eqref{Eq2} display the same universal
behavior of a classical systems in $d+1$ dimension where 
the interaction is LR in $d$ directions and
SR in the remaining Trotter dimension \cite{Dutta2001}. 

Thus a convenient ansatz for the effective action of an $O(N)$ quantum rotor model is
\begin{align}
\label{Eq5}
\Gamma_{k}=\int\,d\tau\,\int\,d^{d}x\{&K_{k}\partial_{\tau}\varphi_{i}\partial_{\tau}\varphi_{i}-Z_{k}\varphi_{i}\Delta^{\frac{\sigma}{2}}\varphi_{i}\nonumber\\
-&Z_{2,k}\varphi_{i}\Delta\varphi_{i}+U_{k}(\rho)\}
\end{align}
where $\Delta$ is the spatial Laplacian in $d$ dimensions, $\tau$ is the 
"Trotter"/imaginary time direction, $\varphi_{i}(x)$ is the $i$-th component 
($i \in \{1,\cdots,N\}$) of the system magnetization density, $\rho\equiv \sum\frac{\varphi_{i}^{2}}{2}$ is the system order parameter, $U_{k}(\rho)$ is the scale dependent effective potential and ($K_{k}$,$Z_{k}$,$Z_{2,k}$) are three scale dependent wave-function renormalization terms. 
In equation \eqref{Eq5} the summation over repeated indexes is intended.

The effective action ansatz \eqref{Eq5} is sufficient to investigate, 
at least at approximate level, the low energy behavior of the models 
described in Hamiltonians \eqref{Eq2} and \eqref{Eq1}. The presence 
of two kinetic terms in the $d$ spatial
directions are necessary to take into account for the competition between 
the LR non analytic momentum term $q^{\sigma}$ in the propagator and 
the usual $q^{2}$ term in the $\sigma\simeq 2$ region.

The frequency and momentum dependence of the propagator at criticality 
are connected by the dynamical critical exponent $z$, 
defined by $\omega\propto q^{z}$. We also expect the 
momentum dependence of the propagator $G(q)$ for large wavelength to obey 
the scaling form
\begin{align}\label{lim}
\lim_{q\to 0}G(q)\propto q^{2-\eta}.
\end{align}
Equation \eqref{lim} defines 
the anomalous dimension $\eta$ as the deviation of the long 
wavelength behavior of the propagator from the standard SR mean field 
behavior $q^{2}$. Finally, as for CPT, the correlation length of the 
quantum system diverges close to
the critical point. Such divergence is power-law, at least 
for standard SSB, and the scaling behavior reads
\begin{align}
\label{Eq7}
\xi\propto \left(\lambda-\lambda_{c}\right)^{-\nu}
\end{align}
where $\xi$ is the correlation length, $\lambda$ is the coupling appearing in Hamiltonian \eqref{Eq2} 
and $\lambda_{c}$ is the critical value of the coupling at which the 
system undergoes SSB. Then equation \eqref{Eq7} can be considered 
the definition of the critical exponent $\nu$.
The universal behavior of any second order QCP can 
be described only in terms of the three exponents $(z,\eta,\nu)$ 
with all the other exponents given by scaling relations \cite{Sachdev2011}. 

\subsection{Local potential approximation}

The lowest order approximation for the ansatz \eqref{Eq5} is 
obtained considering only
the actions parametrized by the effective potential $U_{k}(\rho)$. Imposing
\begin{align}
&Z_{k}=K_{k}=1,\nonumber\\
&Z_{2,k}=0,
\end{align}
it is possible to derive the flow equation in the local potential 
approximation (LPA). The result is equivalent
to the one obtained in the case of classical anisotropic 
$O(N)$ spin system \cite{Defenu2016}. One gets
\begin{align}
\label{Eq9}
\partial_{t}U_{k}(\rho)&=(d+z)\,\tilde{U}_{k}(\tilde{\rho})-(d+z-\sigma)\,\tilde{\rho}\,\tilde{U}^{(1)}(\tilde{\rho})\nonumber\\
&-\frac{1}{1+\tilde{U}^{(1)}(\tilde{\rho})+2\,\tilde{\rho}\tilde{U}^{(2)}(\tilde{\rho})}-\frac{N-1}{1+\tilde{U}^{(1)}(\tilde{\rho})}.
\end{align}
The superscript $\sim$ indicates rescaled quantities, 
which are defined in order to ensure scale invariance 
$\Gamma=\tilde{\Gamma}$ of the effective action at the critical 
point
\begin{align}
U(\rho)&=k^{D_{U}}\tilde{U}(\tilde{\rho})\\
\varphi&=k^{D_{\varphi}}\tilde{\varphi},
\end{align}
with
\begin{align}
D_{U}&=d+z\\
D_{\varphi}&=d+z-\sigma.
\end{align}

At LPA level the kinetic term is fixed and does not renormalize, 
thus anomalous dimension effects 
vanish in both the frequency and momentum sectors. 
Then both the anomalous dimension  
and the dynamical critical exponent attain their mean field values
\begin{align}
\eta&=2-\sigma,\label{eta_mf}\\
z&=\frac{\sigma}{2}.\label{z_mf}
\end{align}
Also equation \eqref{Eq9} is equivalent to the flow for the effective potential in the LPA approximation
for a classical SR spin system \cite{Codello2013,Codello2015} 
in an effective fractional dimension 
\begin{align}
d''_{{\rm SR}}=\frac{2(d+z)}{\sigma},
\end{align}
with the latter equivalence does not account 
for anomalous dimension effects. Substituting
the mean field result for the exponent $z$ 
one obtains
\begin{align}
d_{{\rm SR}}'=\frac{2d}{\sigma}+1,
\label{Eq17}
\end{align}
which is exact in the vanishing anomalous dimension cases \cite{Joyce1966}. 
Similar effective dimension relations have already been introduced in 
literature for the diluted model \cite{Dutta2003}, where the 
random disorder is introduced to simulate LR effects. The critical exponents of
the LR model can be obtained from the ones of the SR model in dimension $d_{\text{SR}}$ via some simple linear transformations and they fulfill all the
usual scaling relations included hyper-scaling, as long as $d<d_{\text{uc}}$.  
We refer to \cite{Defenu2015} for a discussion on the reliability of the
effective dimension, see as well \cite{Behan2017a}. 

We are then in position to predict the upper critical dimension, 
which is obtained from the condition $d_{{\rm SR}}\geq 4$, in such 
a way that
\begin{align}
\label{Eq19}
d_{{\rm uc}}=\frac{3}{2}\sigma, 
\end{align}
in agreement with the one obtained by relevance arguments in \cite{Dutta2001}.
For $d>d_{uc}$ the system undergoes SSB with mean field 
critical exponents given in 
equations \eqref{eta_mf} and \eqref{z_mf}, with 
\begin{align}
\nu&=\sigma^{-1}.\label{nu_mf}
\end{align}

For continuous symmetries in the SR interacting case the anomalous dimension
vanishes also at the lower critical dimension as ensured by the MWT. 
Thus one can again employ the effective dimension approach to obtain
\begin{align}
\label{Eq20}
d_{{\rm lc}}=\frac{\sigma}{2}.
\end{align}
It is worth noting that, while relation \eqref{Eq19} is valid for all $N$, 
included the $N=1$ Ising case, 
the expression \eqref{Eq20} is limited to continuos symmetries $N\geq 2$.

Summarizing, 
the final picture emerging from LPA approximation is rather simple. 
Equation \eqref{Eq9} has, in general,
two fixed point solutions, one is the Gaussian fixed point, 
which represents the mean field universality,
characterized by the critical exponents in equations 
\eqref{eta_mf}, \eqref{z_mf} and \eqref{nu_mf}.
The other solution has finite renormalized mass 
$\tilde{U}^{(1)}(0)\neq 0$ and represents the
interacting Wilson-Fisher (WF) universality.

For $d\geq d_{{\rm uc}}$ the Gaussian universality 
is the attractive one and the quantum LR system
will display SSB with mean field exponents. 
On the other hand for $d<d_{{\rm uc}}$ the attractive 
fixed point is the WF one  and no analytic expression
for the critical exponents is known. 
Finally for $d\leq d_{{\rm lc}}$ no SSB is possible for continuous symmetries
and the system will have a single phase in the $N\geq 2$ case. 

At the border region $\sigma\simeq 2$ the LR and SR momentum terms 
are competing in the propagator and anomalous dimension effects become increasingly 
important \cite{Codello2015}. In this section we discarded 
anomalous dimension effects and the threshold values at which LR interactions 
become irrelevant
with respect to SR ones is $\sigma_{*}=2$ as can be detected 
by mean field arguments \cite{Dutta2001}. However, it seems by now 
established that the actual boundary between LR and SR critical behavior 
is given
by Sak's expression $\sigma_{*}=2-\eta_{{\rm SR}}$ 
\cite{Sak1973}, where $\eta_{{\rm SR}}$ is the anomalous
dimension of the system in the $\sigma=2$ case. In order 
to investigate this effect we shall rely on the complete
ansatz \eqref{Eq5}. 

\subsection{Anomalous dimension effects}

In order to introduce anomalous dimension effects it is necessary
to consider the flow of the kinetic sectors. In the simplified approach
considered in this paper it is sufficient to compute the running of
the coefficients $Z_{k}$, $K_{k}$ and $Z_{2,k}$. 
The flow for the LR wave-function $Z_{k}$ is given by 
\begin{align}
\label{Eq21}
\partial_{t}Z_{k}=\left(2-\sigma-\eta\right)Z_{k},
\end{align}
where the anomalous dimension $\eta$ is defined with respect to the SR term
\begin{align}
\eta=\frac{\partial_{t}Z_{2,k}}{Z_{2,k}}.
\end{align}

At the fixed point the running of the couplings should vanish
and thus we have only two possibilities, either
\begin{align}
\label{Eq22}
\eta=2-\sigma
\end{align}
or
\begin{align}
\label{Eq23}
\lim_{k\to 0}Z_{k}=0.
\end{align}
The latter result will imply that LR interactions are irrelevant
and do not influence the critical behavior. Latter analysis 
is consistent with Sak's result. Indeed for $\sigma<\sigma_{*}$
LR interactions are relevant and $\eta=2-\sigma$, while for
$\sigma\geq\sigma_{*}$ the LR term vanishes at criticality and the system
recovers SR universal behavior $\eta=\eta_{{\rm SR}}$.

Therefore the boundary value $\sigma_{*}$ is obtained when the 
coefficient in the r.h.s of equation \eqref{Eq21} 
vanishes $2-\sigma_{*}-\eta=0$
and $\eta$ attains its SR value $\eta_{{\rm SR}}$:
\begin{align}
\label{Eq24}
\sigma_{*}=2-\eta_{{\rm SR}}.
\end{align}
For $\sigma\geq\sigma_{*}$ the effective action of the system 
at the fixed point becomes isotropic and the dynamical critical exponent 
is $z=1$. Then, according
to quantum to classical correspondance \cite{Mussardo2010,Sachdev2011} 
the QCP of the SR system is in the same universality of the CPT 
of its classical analogous in $d_{{\rm SR}}=d+1$.

\begin{figure*}[ht!]
\subfigure[\large\,\,\, $N<2$]{\label{Fig1a}\includegraphics[width=.45\textwidth]{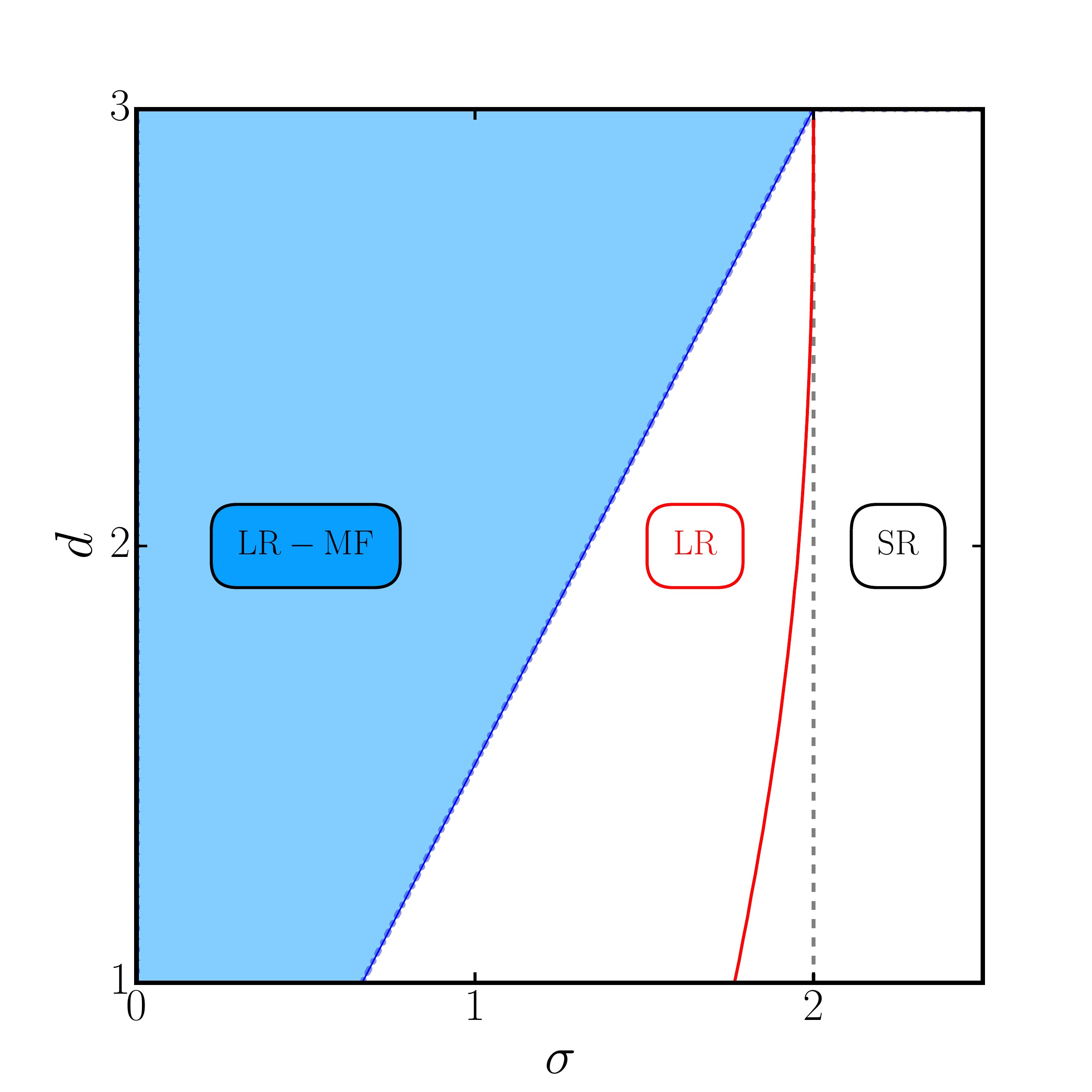}\label{Fig1b}}
\subfigure[\large\,\,\, $N\geq 2$]{
\includegraphics[width=.45\textwidth]{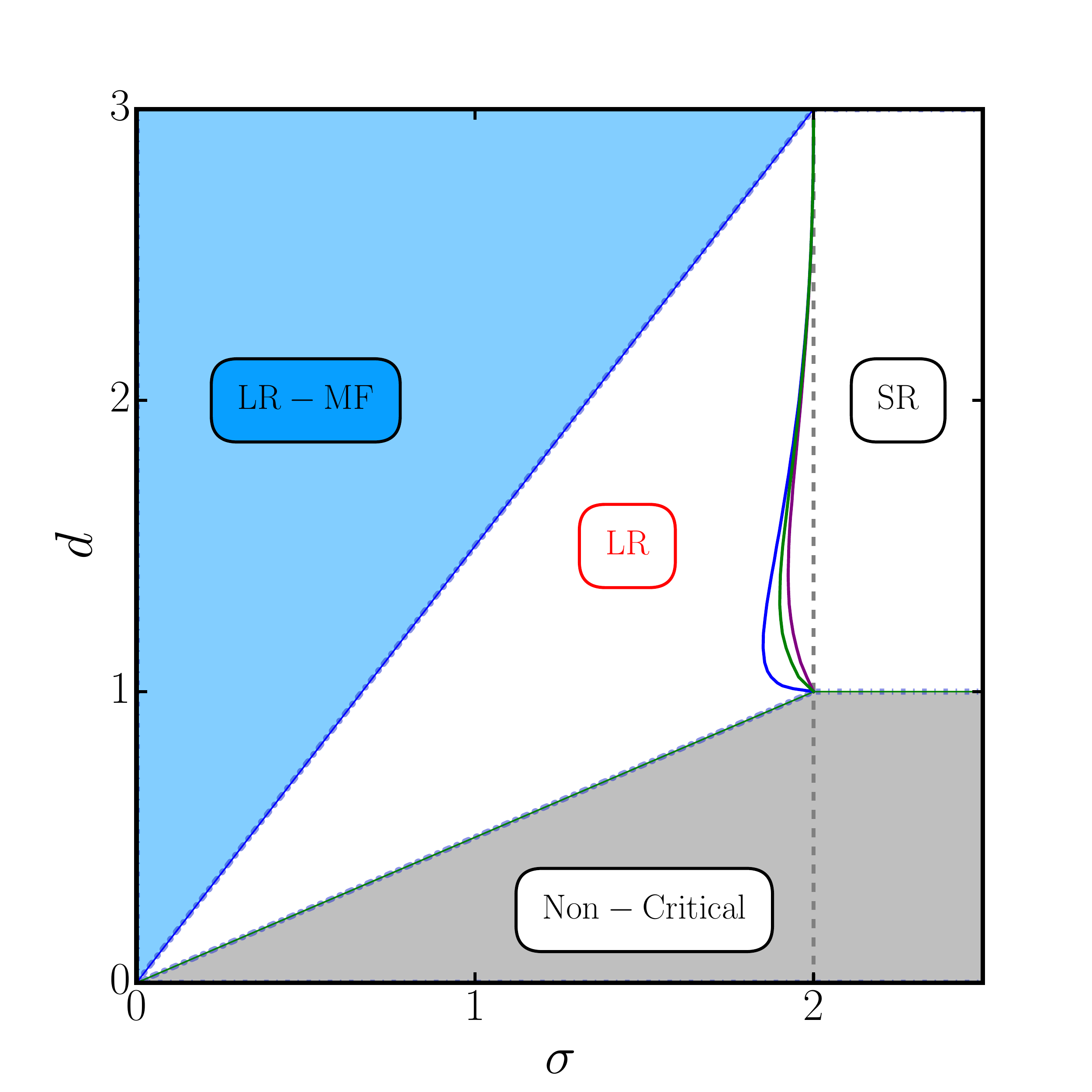}\label{Fig1b}}
\caption{Phase diagram of quantum LR rotor models for discrete, panel (a), 
and continuous symmetries, panel (b).
The cyan shaded area represents the mean field validity region, 
while the gray shaded area is the single phase region. 
The mean field boundary $\sigma_{*}=2$ is represented by a gray dashed line, 
renormalized boundaries are solid
lines in red, blue, green and purple for $N=1,2,3,4$ respectively.}
\label{Fig1}
\end{figure*}

The phase diagram of a quantum LR rotor model in the $(d,\sigma)$ plane 
is reported in figure \ref{Fig1}. 
In the discrete symmetry case, figure \ref{Fig1a}, the system 
has a mean field validity region, indicated
by the cyan shaded area, where the exponents are the ones 
obtained by mean field approximation. Otherwise
for $d<d_{{\rm uc}}$ the system has peculiar non mean field exponents. The boundary
$\sigma_{*}$ after which the system recovers SR behavior 
is indicated by the red solid line, the line has been
computed using expression \eqref{Eq24} with the 
expression for $\eta_{{\rm SR}}$ obtained by ansatz \eqref{Eq5}
without any non analytic terms. For discrete symmetries SSB appears also
in $d=1$.

For continuous symmetries, figure \ref{Fig1b}, 
the scenario is the same as for discrete 
symmetries, except that, due to the MWT, the $\sigma_{*}$ 
boundary recovers its mean field value not only at $d=d_{{\rm uc}}$
but also for $d=d_{{\rm lc}}=1$ at $\sigma_{*}$. For $d\leq d_{{\rm lc}}$, 
gray shaded area, the system displays only a single phase.

\section{The critical exponents}\label{sec:4}

As shown in the previous section the system has a phase 
diagram similar to the one already depicted for a classical
LR spin systems \cite{Angelini2014,Defenu2015} with non trivial
LR behavior occurring only for $\sigma<\sigma_{*}$ and $d<d_{{\rm uc}}$.
In this region no analytical expressions are possible for the critical exponents
and one shall resort to numerical approximated techniques.

In order to compute the critical exponents in the non trivial region one can
apply the same procedure as done in \cite{Defenu2016} for anisotropic LR
spin systems. Again we consider the ansatz \eqref{Eq5} in the case of vanishing
analytic corrections $Z_{2,k}=0$ in the momentum sector. Indeed the latter 
term is relevant only close to the boundary $\sigma_{*}$  and it has been shown
in \cite{Defenu2015,Defenu2016} to produce only minor corrections 
even for $\sigma\simeq\sigma_{*}$

The flow of the effective potential for vanishing analytic 
correction $Z_{2,k}=0$ reads
\begin{equation}
\label{Eq25}
\begin{split}
\partial_t \bar{U}_{k}= &(d+z)\bar{U}_{k}(\bar{\rho})-(d+z-\sigma)\bar{\rho}\,\bar{U}'_{k}(\bar{\rho})\\
&- \frac{\sigma}{2}(N-1)\frac{1-\frac{\eta_{\tau} z}{3\sigma+2d}}{1+\bar{U}'_{k}(\bar{\rho})}
-\frac{\sigma}{2}\frac{1-\frac{\eta_{\tau} z}{3\sigma+2d}}{1+\bar{U}'_{k}(\bar{\rho})+2\bar{\rho}\,\bar{U}''_{k}(\bar{\rho})},
\end{split}
\end{equation} 
where the anomalous dimension $\eta_{\tau}$ has been introduced. This quantity introduces a correction in the frequency dependence of the 
system propagator
\begin{align}
\label{Eq26}
\lim_{\omega\to 0}G(\omega,1)^{-1}\propto \omega^{2-\eta_{\tau}}.
\end{align}
The expression for the frequency anomalous dimension is readily obtained by the flow of $K_{k}$ 
\begin{align}
\label{Eq27}
\eta_{\tau}=\frac{f(\tilde{\rho}_{0},\tilde{U}^{(2)}(\tilde{\rho}_{0}))(3\sigma+2 d)}{d+(3\sigma+d)(1+f(\tilde{\rho}_{0},\tilde{U}^{(2)}(\tilde{\rho}_{0})))},
\end{align}
where the function $f(\tilde{\rho}_{0},\tilde{U}^{(2)}(\tilde{\rho}_{0}))$ 
is the expression for the spatial anomalous dimension of the correspondent SR range 
$O(N)$ model
\begin{equation}
\label{Eta_SR}
f(\tilde{\rho}_{0},\tilde{U}^{(2)}(\tilde{\rho}_{0}))=\frac{4\tilde{\rho}_{0}\tilde{U}^{(2)}(\tilde{\rho}_{0})^{2}}{(1+2\tilde{\rho}_{0}\tilde{U}^{(2)}(\tilde{\rho}_{0}))^{2}}
\end{equation}
as is found in \cite{Codello2013} after rescaling an unessential 
geometric coefficient.

Merging the definition of the dynamical critical exponent $z$ with 
the definition of
$\eta_{\tau}$ in equation \eqref{Eq26} it is readily obtained that 
\begin{align}
z=\frac{\sigma}{2-\eta_{\tau}},
\end{align}
which has to be compared with the mean field expression \eqref{z_mf}.

Eq. \eqref{Eq25} allows for an higher order effective dimension relations
which connects the universal behavior of a $d$ dimensional quantum LR system to 
its classical SR counterpart in $d_{\text{SR}}$ dimensions with
\begin{align}
\label{Eq31}
d_{\text{SR}}=(2-\eta_{\text{SR}})\frac{(d+z)}{\sigma},
\end{align}
where $\eta_{\text{SR}}$ is the anomalous dimension of the SR system in dimension $d_{\text{SR}}$.
Such relation can be easily demonstrated using the procedure depicted in 
\cite{Defenu2015} to relate Eq. \eqref{Eq25} to the potential flow shown in \cite{Codello2015}. However the exact demonstration only holds if we neglect
the coefficients proportional to $\eta_{\tau}$ appearing in the second line
of equation \eqref{Eq25}. These coefficients are small regulator dependent 
quantities which produce $O(\eta_{\tau}^{2})$ contributions to the universal 
quantities. It has been verified that neglecting such coefficients produces numerical errors  below 1\% in the solution of flow equations like \eqref{Eq25} \cite{Berges2002,Defenu2015}.

\begin{figure*}[ht!]
\subfigure[\large\,\,\, $d=1$]{\includegraphics[width=.45\textwidth]{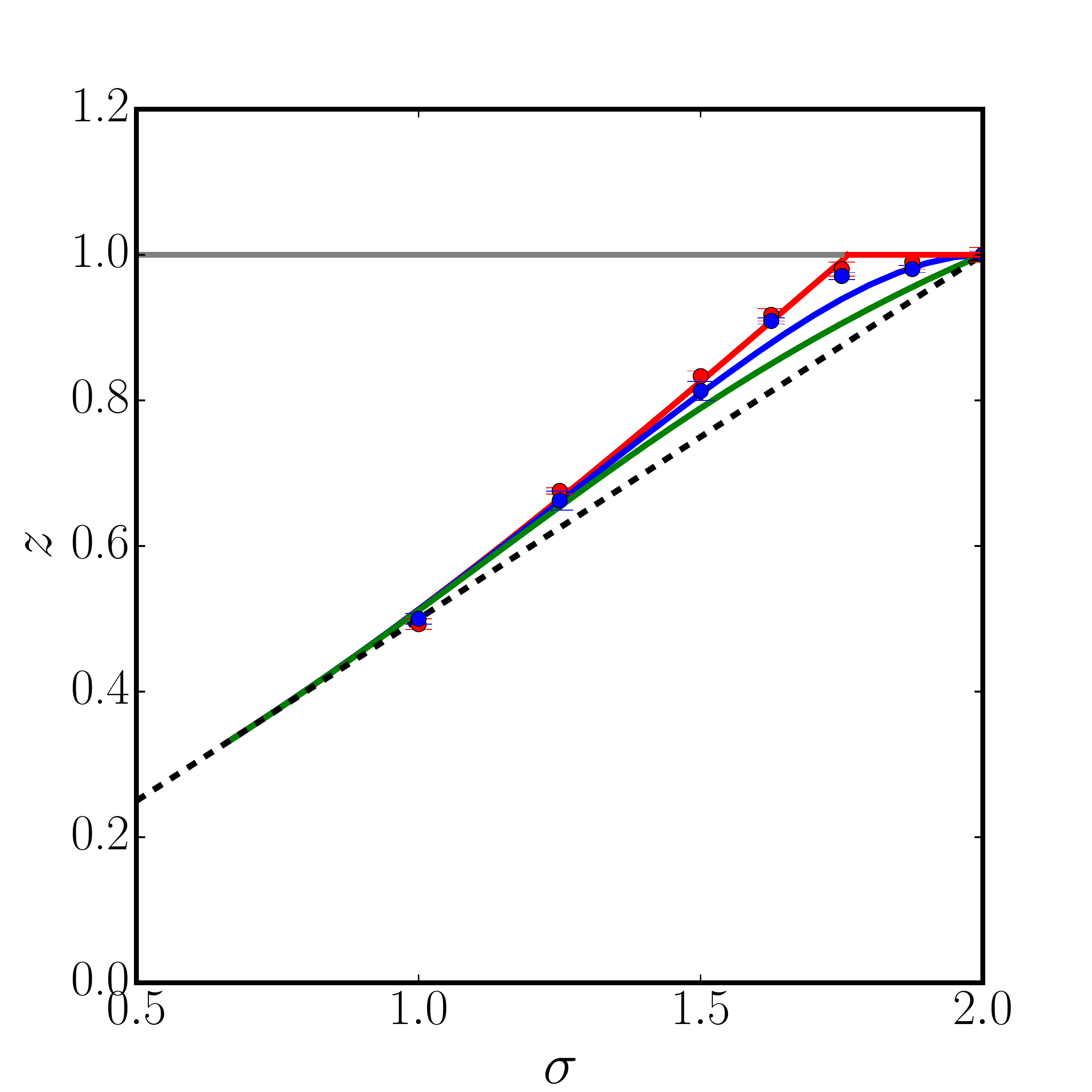}\label{Fig2a}}
\subfigure[\large\,\,\, $d=2$]{
\includegraphics[width=.45\textwidth]{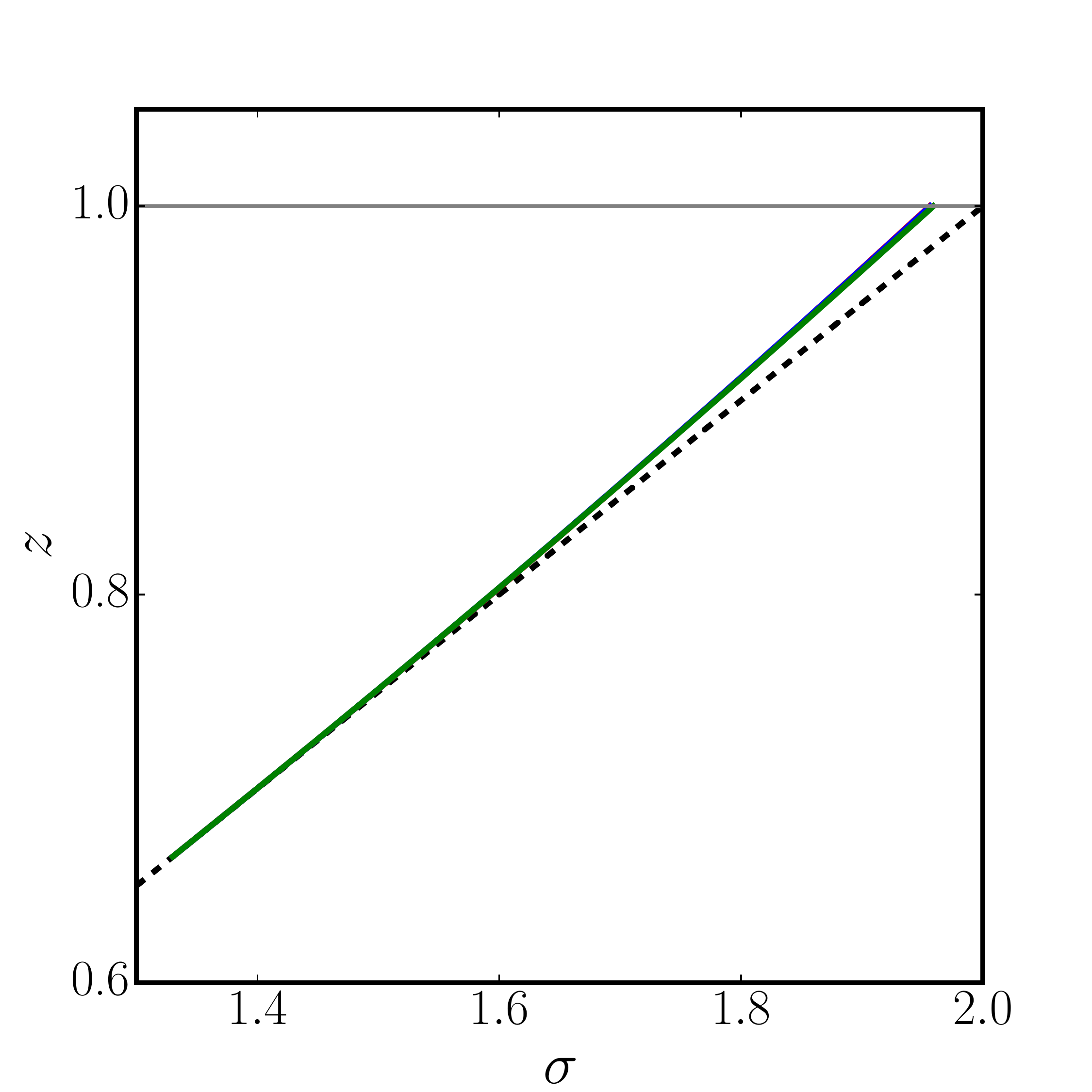}\label{Fig2b}}
\caption{The dynamical critical exponent $z$ as a function of 
$\sigma$ in $d=1,2$, panel (a) and (b) respectively. The red, blue and green 
curves are for $N=1,2,3$ respectively, the mean field result is shown as a 
gray dashed line. 
The (blue and red) dots represent MC data for $d=1$ derived for 
a spin system coupled to a bosonic 
bath with spectral density proportional to $\omega^{\sigma}$, data taken from 
\cite{Sperstad2012}, with the red dots for the Ising model 
(in good agreement with our predictions) and the 
blue dots for the $O(2)$ model.}
\label{Fig2}
\end{figure*}

In figure \ref{Fig2} the dynamical critical exponent 
$z$ as a function of the decay exponent $\sigma$ is shown in the $d=1,2$, 
panel (a) and (b) respectively. The data have been obtained solving 
the expression for the fixed point effective potential, 
equation \eqref{Eq25} with the l.h.s posed to zero, 
and equation \eqref{Eq27} in a self consistent cycle.
The dynamical critical exponent $z$ attains its mean field value 
for $\sigma<\frac{2}{3}d$, while it becomes a non trivial curve in 
the non Gaussian region. It should be noted that correlation effects 
always increase the value of the dynamical critical exponent 
with respect to the mean field prediction. However in the continuos symmetry 
cases $N=2,3$, for $d=1$, due to the MWT the curves bend 
down at some finite value $\sigma$ and the dynamical critical exponent 
$z$ recovers its SR value only at $\sigma_{*}=2$. 
On the other hand in the $d=2$ case the Ising model curve (red solid line) is monotonically increasing and met the SR line 
(gray solid line in figure \ref{Fig2}) at the value $\sigma_{*}<2$.

The dynamical critical exponent values obtained by the present approach 
for the Ising model, red solid line in \ref{Fig1a}, are in very good 
agreement with MC simulation on an Ising spin system coupled 
to an anomalous Bosonic bath, which lies in the same universality, red dots in \ref{Fig1a}. The agreement is poorer in the $O(2)$ case, blue solid line theoretical values and blue dots MC simulation.  The MC
data are taken from reference \cite{Sperstad2012}.  

The lower accuracy found for the $N=2$ can be due to the presence of the 
BKT mechanism for this model. In passing we note that the MC points seems 
to provide a value $\sigma_{*}=2-\eta$ with $\eta=\frac{1}{4}$ 
even for the $N=2$, which is consistent with the BKT scenario. 

\begin{figure*}[ht!]
\subfigure[\large\,\,\, $d=1$]{\includegraphics[width=.49\textwidth]{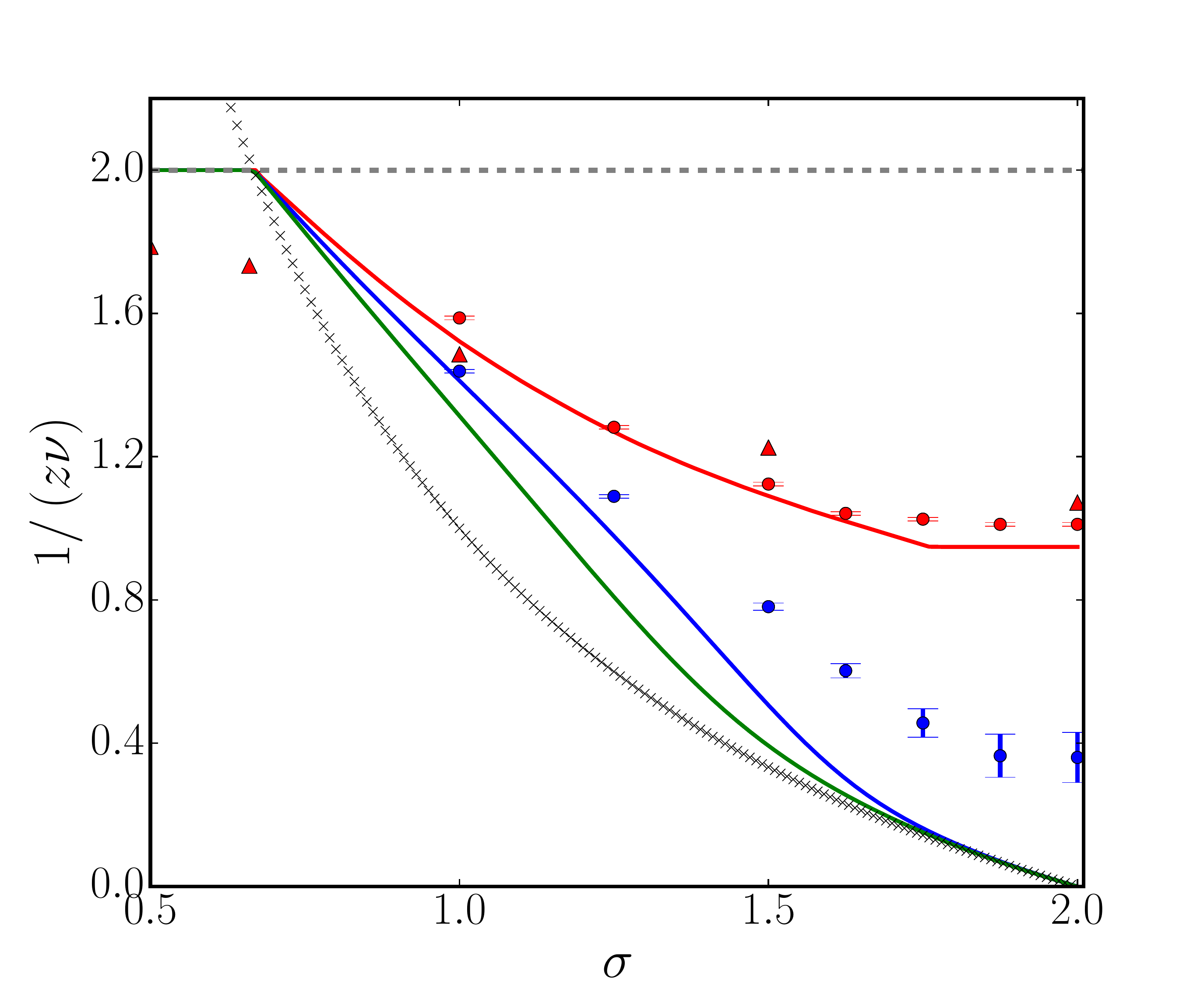}\label{Fig3a}}
\subfigure[\large\,\,\, $d=2$]{\includegraphics[width=.49\textwidth]{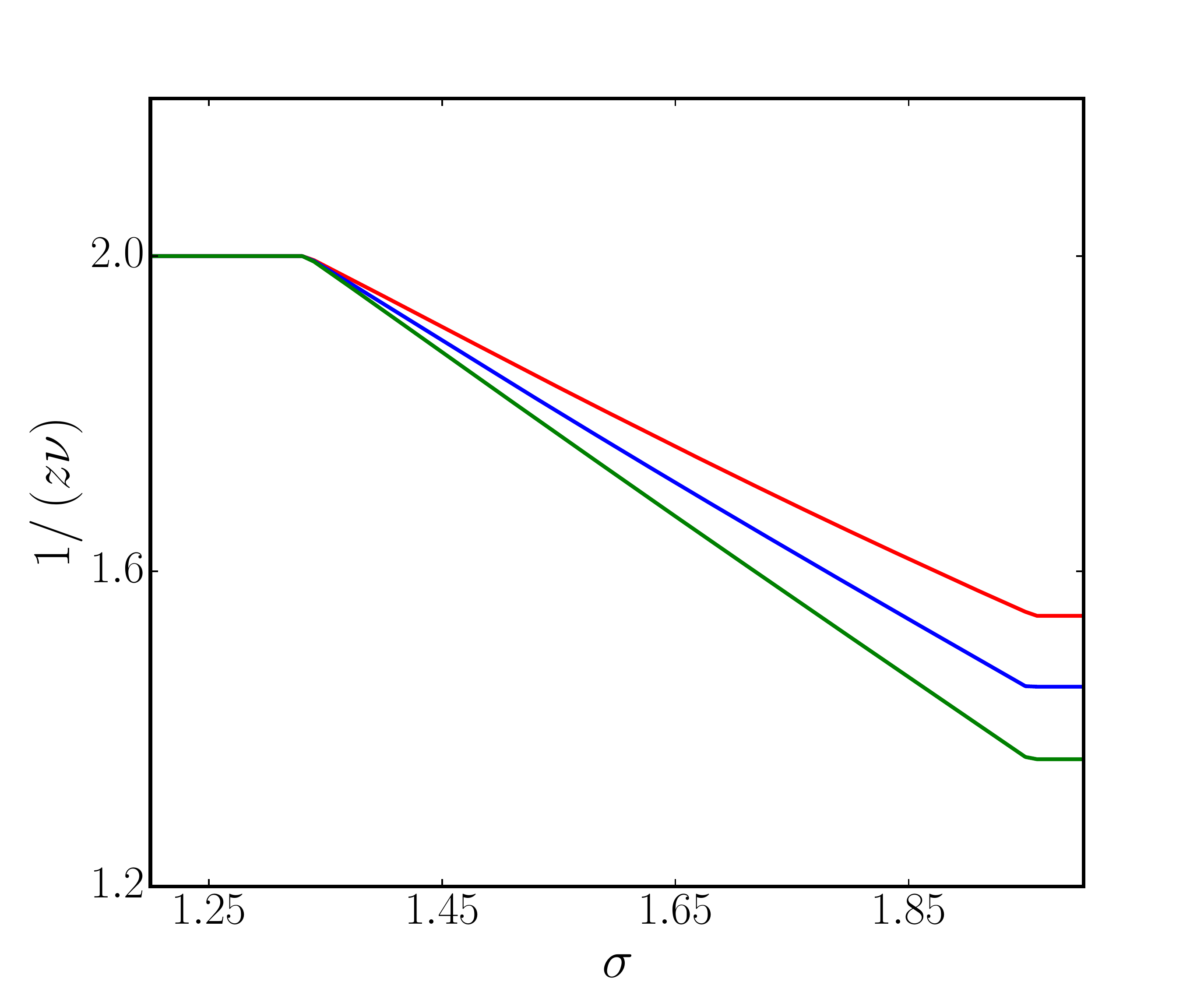}\label{Fig3b}}
\caption{The $(z\nu)^{-1}$ as a function of $\sigma$ in $d=1,2$, panel 
(a) and (b) respectively. The red, blue and green curves are for $N=1,2,3$ 
respectively, the mean field result is shown as a gray dashed line. 
Again dots represent MC data for the Ising (red) and the $O(2)$ model (blue) 
in $d=1$, data from reference \cite{Sperstad2012}. {The red triangles represent
estimation obtained with perturbative continuous unitary transformation taken from \cite{Fey2016}. These data however cannot reproduce the expected values in the mean field 
region and seem unable to correctly account for the expected value of the SR boundary $\sigma_{*}=2-\eta_{\text{SR}}$.  Our results for the correlation length exponent $\nu$ have been obtained using the effective dimension relation \eqref{Eq31} to map
the values of $\nu$ for classical SR systems found in \cite{Codello2015}
to the quantum LR case, see the procedure outlined in \cite{Defenu2015}.
It should be noted that for the SR Ising case it is necessary to employ two different
definitions of anomalous dimensions in order to obtain reliable results , see \cite{Codello2013,Codello2015,Defenu2015}}. In panel (a) 
leading order terms of $\varepsilon=(d_{uc}-d)$ and 
$\tilde{\varepsilon}=(d-d_{lc})$ expansion are shown as gray dot-dashed and 
gray crosses respectively.}
\label{Fig3}
\end{figure*}

The presented computation, despite having the merit to be valid 
for any $d$, $\sigma$ and $N$, is anyway 
not able to fully capture the effects of the topological 
phase transition and no qualitative difference is found between the $N>2$ 
and the $N=2$ for $\sigma=2$ in $d=1$.
However, close to $\sigma=2$ for $N\geq 2$ it appears that 
MC simulations may be as well plagued by finite size effects
as it will become evident discussing the results of figure \ref{Fig3}, 
since they do not reproduce the expected limiting behavior 
$\sigma \to 2$ for the correlation length exponent, {see  also the discussion in} \cite{Sperstad2012}. 
On the other hand our calculation of the quantity $(z\nu)^{-1}$, 
see figure \ref{Fig3}, is fully consistent with the expected exact 
behavior coming from $d-2$ expansions \cite{Brezin1976} and the 
discrepancy found between the theoretical prediction and the MC data for 
$z$ in the $N=2$ case is perhaps due to finite size errors of 
the MC data rather than to unexpected BKT effects for $\sigma<2$.

In order to completely characterize the critical behavior of 
quantum LR models it is necessary to derive the values
for the correlation length critical exponent in the non trivial region. 
In order to accomplish this task it is sufficient 
to make a linear perturbation of the fixed point potential with 
the form $\tilde{U}(\tilde{\rho})=\tilde{U}_{*}(\tilde{\rho})
+u(\tilde{\rho})e^{y_{t}\,t}$, where the $*$ denotes the any scaling 
solution of equation \eqref{Eq25}. 

Employing the technique already introduced in \cite{Codello2015,Morris1998a} 
it is possible to obtain the discrete spectrum of the eigenvalues $y_{t}$, 
which describe the scaling behavior of any physical perturbation 
of the fixed point effective action.
As it is well known for standard second order phase transitions 
only one relevant perturbation of the critical Ginzburg-Landau 
free energy 
exist, which either drives the system to its high or low temperature phases. 
Then the eigenvalues spectrum will contain only one positive solution, 
namely the inverse correlation exponent ${\rm max}\,\, y_{t}=\nu^{-1}$.

{Another equivalent way to obtain the correlation length exponent $\nu$ is to
take advantage of the effective dimension relation \eqref{Eq31} in order to
map the values of $\nu$ in the classical SR case to the quantum LR case via 
the relation
\begin{align}
\label{Eq32}
\nu=\frac{2-\eta_{\text{SR}}}{\sigma}\nu_{\text{SR}},
\end{align}
where $\nu_{\text{SR}}$ is the anomalous dimension of the SR system
in dimension $d_{\text{SR}}$. This procedure is equivalent to solving 
the stability spectrum of equation \eqref{Eq25}
once the regulator dependent coefficients in the non-linear part
have been neglected \cite{Defenu2015}}. In figure \ref{Fig3} 
the results for the exponent $(z\nu)^{-1}$ are reported in the case of a 
quantum LR rotor model in dimension $d=1$ and $d=2$ panel \ref{Fig3a} 
and \ref{Fig3b} respectively. 
The results for the Ising criticality (red solid line) 
are in good agreement with the numerical findings (red dots). 

On the other hand the $N=2$ case 
[blue solid line in figure \ref{Fig3a}] is, once again, in disagreement 
with numerical findings (blue dots). In this case, it appears that 
MC simulations are not completely reliable since they do not 
reproduce the exact behavior of the correlation length exponent in the 
$\sigma \to 2$ limit, where it is expected to diverge, {see also the discussion
in \cite{Sperstad2012}}. Such divergence is 
in agreement with the BKT behavior, where exponential 
divergence of the correlation length is found, as 
it can be derived exactly generalizing the $2+\tilde{\varepsilon}$ 
expansion technique of the SR non linear $\sigma$-model in 
reference \cite{Brezin1976} to the present case. 

Latter generalization is more readily obtained by means of 
effective dimension approach. One should consider equation (21) of
reference \cite{Brezin1976} at lowest order in $\tilde{\varepsilon}=d-2$
\begin{align}
\nu_{SR}^{-1}=(d-2)+O(\tilde{\varepsilon}^{2})
\end{align} 
employing the relation \eqref{Eq31}
with the effective dimension $d_{\rm SR}$ given by equation 
\eqref{Eq17}, in the limit of vanishing anomalous dimension.
The result is
\begin{align}
\nu\simeq 1-\frac{\sigma}{2}
\end{align}
which when multiplied by $z^{-1}$ gives
\begin{align}
\lim_{\sigma\to2}(z\nu)^{-1}=\frac{2}{\sigma}-1,\quad \forall\,\, N\geq2.
\end{align}
It should be noted that the derivation above does not consider anomalous dimension
effects which we expect to be sub-leading with respect to the main contribution
to $\nu$.
The latter result (black dot dashed line in \ref{Fig3a}) 
is in perfect agreement with our findings, 
but it cannot be reproduced by MC simulations possibly 
due to finite size effects close to the $\nu^{-1}\simeq 0$ region.

While we expect numerical simulations to be very accurate only for $\sigma$ 
well below $2$, it is also found that our results for $(z\nu)^{-1}$ should 
be slightly underestimated in the intermediate $\sigma$ region, 
even if reproducing all the limiting behaviors. This is in agreement
with previous investigation on classical LR and SR O(N) models \cite{Codello2013,Codello2015,Defenu2015}, where the FRG results for the correlation length exponent 
is found to slightly overestimate high precision numerical estimation in the
whole dimension range. Here the same
effect is present as a function of the exponent $\sigma$.
Thus it is fair to conclude 
that the exact correlation length exponent curve for the $N=2$ case in the 
$d=1$ case should be located in between the theoretical 
curve and the reported numerical estimates.
Apart from the $N=2$ case, which is rather special exhibiting the BKT 
transition, we expect our findings to be very accurate for all $N>2$,{ since
for large components number ($N\to \infty$) the ansatz \eqref{Eq5} becomes exact.} 

\section{The Spherical model case}

In the infinite components limit ($N\to \infty$) the classical $O(N)$ model lies in the same universality of the spherical model, which is exactly solvable. Within  the classical analogue of ansatz \eqref{Eq5} the exact expressions for the critical exponents can be recovered \cite{Codello2013,Codello2015}. Such property also remains valid in the quantum case
\cite{Nieuwenhuizen1995,Vojta1996} leading to 
\begin{align}
z &=\frac{\sigma}{2},\\
\nu &=\frac{2}{2d-\sigma} \label{Eq36}.
\end{align}
These expressions, obtained from the exact solution of the flow equations in the $N\to\infty$ limit, give the exact values of the critical exponents for the quantum spherical model \cite{Vojta1996}. 

It should be noted that the same expression for the critical exponents could be obtained via the effective dimension relations. Indeed in the classical SR $O(N)$ models we have $\nu=(d_{\text{SR}}-2)^{-1}$ and $\eta=0$, then employing the effective dimension relation, equation \eqref{Eq17}, and multiplying  by $2/\sigma$ we immediately get \eqref{Eq36}. Thus the effective dimension relation is exact in the $N\to\infty$ limit, as
it is expected from the classical spherical model case \cite{Joyce1966}.

\section{Conclusions}\label{sec:5}

The results in figure \ref{Fig3} completes the necessary information to 
derive the scaling behavior of all the thermodynamic quantities of quantum 
LR rotor models. Indeed we determined both the dynamical critical 
exponent $z$ and the inverse correlation length exponent $\nu^{-1}$ 
from which all the critical exponents can be obtained via the scaling relations. It should be noted that, in principle, 
one should also know the values for the anomalous dimension $\eta$, 
which however in the LR system is found to be always given by the mean field 
result $\eta=2-\sigma$. In table \ref{tab1} the values of the critical exponents
for various values of $N,d$ and $\sigma$ are reported.
\begin{table}[ht]
 \caption{Critical exponents for quantum LR $O(N)$ models for various 
values of $N,d$ and $\sigma$. The rational 
quantities are the exact results, in the mean-field validity region. The 
 numerical values are obtained by self consistent solutions of the flow equations with numerical precision $\pm 0.002$.}
\centering 
\begin{tabular}{|c c | c c | c c|} 
\cline{3-6}
\multicolumn{2}{c|}{} & \multicolumn{2}{c|}{$d=2$}&\multicolumn{2}{c|}{$d=3$}\\
\hline 
$N$ & $\sigma$ & $z$ & $1/\nu$ &$z$ & $1/\nu$ \\ [.5ex] 
\hline 
1 & 2/3 & 1/3 & 2/3 & 1/3 & 2/3 \\ 
2 & 2/3 & 1/3 & 2/3 & 1/3 & 2/3\\
3 & 2/3 & 1/3 & 2/3 & 1/3 & 2/3\\
1 & 4/5 & 0.402 & 0.720 & 2/5 & 4/5\\ 
2 & 4/5 & 0.402 & 0.703 & 2/5 & 4/5\\
3 & 4/5 & 0.402 & 0.689 & 2/5 & 4/5\\
1 & 1 & 0.513 & 0.780 & 1/2 & 1\\ 
2 & 1 & 0.513 & 0.706 & 1/2 & 1\\
3 & 1 & 0.512 & 0.657 & 1/2 & 1\\
1 & 6/5 & 0.633 & 0.832 & 3/5 & 6/5\\ 
2 & 6/5 & 0.630 & 0.641 & 3/5 & 6/5\\
3 & 6/5 & 0.626 & 0.545 & 3/5 & 6/5\\
1 & 7/5 & 0.761 & 0.878 & 0.700 & 1.365\\ 
2 & 7/5 & 0.751 & 0.487 & 0.700 & 1.357\\
3 & 7/5 & 0.737 & 0.376 & 0.700 & 1.351\\
1 & 8/5 & 0.892 & 0.921 & 0.803 & 1.442\\ 
2 & 8/5 & 0.866 & 0.269 & 0.803 & 1.405\\
3 & 8/5 & 0.839 & 0.224 & 0.803 & 1.377\\
1 & 9/5 & 1.000 & 0.948 & 0.911 & 1.504\\ 
2 & 9/5 & 0.959 & 0.110 & 0.912 & 1.422\\
3 & 9/5 & 0.926 & 0.105 & 0.911 & 1.364\\
1 & 2 & 1.000 & 0.948 & 1.000 & 1.543\\ 
2 & 2 & 1.000 & 0.000 & 1.000 & 1.417\\
3 & 2 & 1.000 & 0.000 & 1.000 & 1.327\\ [1ex] 
\hline 
\end{tabular}
\label{tab1} 
\end{table}

Our motivations for the study presented in this paper were two-fold: on 
one side the progress in the control of 
atomic, molecular and optical systems made it possible to experimentally 
implement quantum long-range (LR) systems with tunable parameters, 
including the range of the interactions. These advancements pave the way 
to the study of the phase diagram and of the criticality quantum LR models, 
calling for the development of a unified treatment of their properties. 
On the other side, this motivates the search for the analogies and 
differences between quantum and classical LR systems, to understand 
what phenomena in classical systems typically due to the long-rangeness 
are also present in quantum systems and what are specifically connected to 
the presence of quantum fluctuations. Finally, a last motivation 
for the present work was to have a compact formalism to study 
 different quantum LR systems at once, allowing also to clarify the 
relation between quantum LR models in $d$ dimensions 
and the corresponding anisotropic classical models in $d+1$ dimensions.

We therefore developed a general description of quantum LR models 
based on renormalization group (RG) techniques and capable to work 
for different dimensions $d$, power-law decay exponents $\sigma$ and 
groups of symmetry. We focused  
on the derivation of universal exponents describing the critical
behavior of power-law decaying interacting $N$ components quantum rotor models.
The $N=1$ case corresponds to a quantum LR Ising model, which
is described by Hamiltonian \eqref{Eq1}, while the general $N\geq 2$ cases, 
described by Hamiltonian \eqref{Eq2}, are quantum generalizations 
of the celebrated $O(N)$ symmetric models.

LR power decaying interactions deeply affect the critical behavior of 
quantum models and their critical dynamics. We remind that the
usual short-range (SR) quantum model with local propagator 
$G^{-1}(q)\propto q^{2}$ has a quantum
critical point at zero temperature which lies in the same universality 
class of the classical
phase transition in $d+1$ dimension \cite{Sachdev2011}. 
As discussed in the 
text, in the LR interacting case the low energy behavior of the
propagator is not analytic in the momentum, being 
$G^{-1}(q)\propto q^{\sigma}$, and 
the quantum field theory
obtained using Trotter decomposition is 
anisotropic in the extended $d+1$ dimensional space. However 
our analysis clearly shows that the problem 
is reduced to the study of a LR classical problem only for 
$\sigma$ large enough, and that this is not the case for $\sigma$ smaller 
than a critical value. For such small values of $\sigma$ 
the correlation functions are genuinely strongly anisotropic in the spatial 
and time coordinates and 
the isotropy is not restored even at the criticality.

The mapping of a quantum SR universality into its 
$d+1$ classical equivalent obtained by Trotter decomposition is exact and 
so it is the relation between the universal 
behavior of a quantum LR model and 
its anisotropic classical equivalent. On the other hand, from
the first line of equation \eqref{Eq25} 
a correspondance emerges between the quantum LR system in dimension
$d$ and its classical analogous in dimension $d+z$. 

\begin{figure*}[ht!]
\includegraphics[width=.45\textwidth]{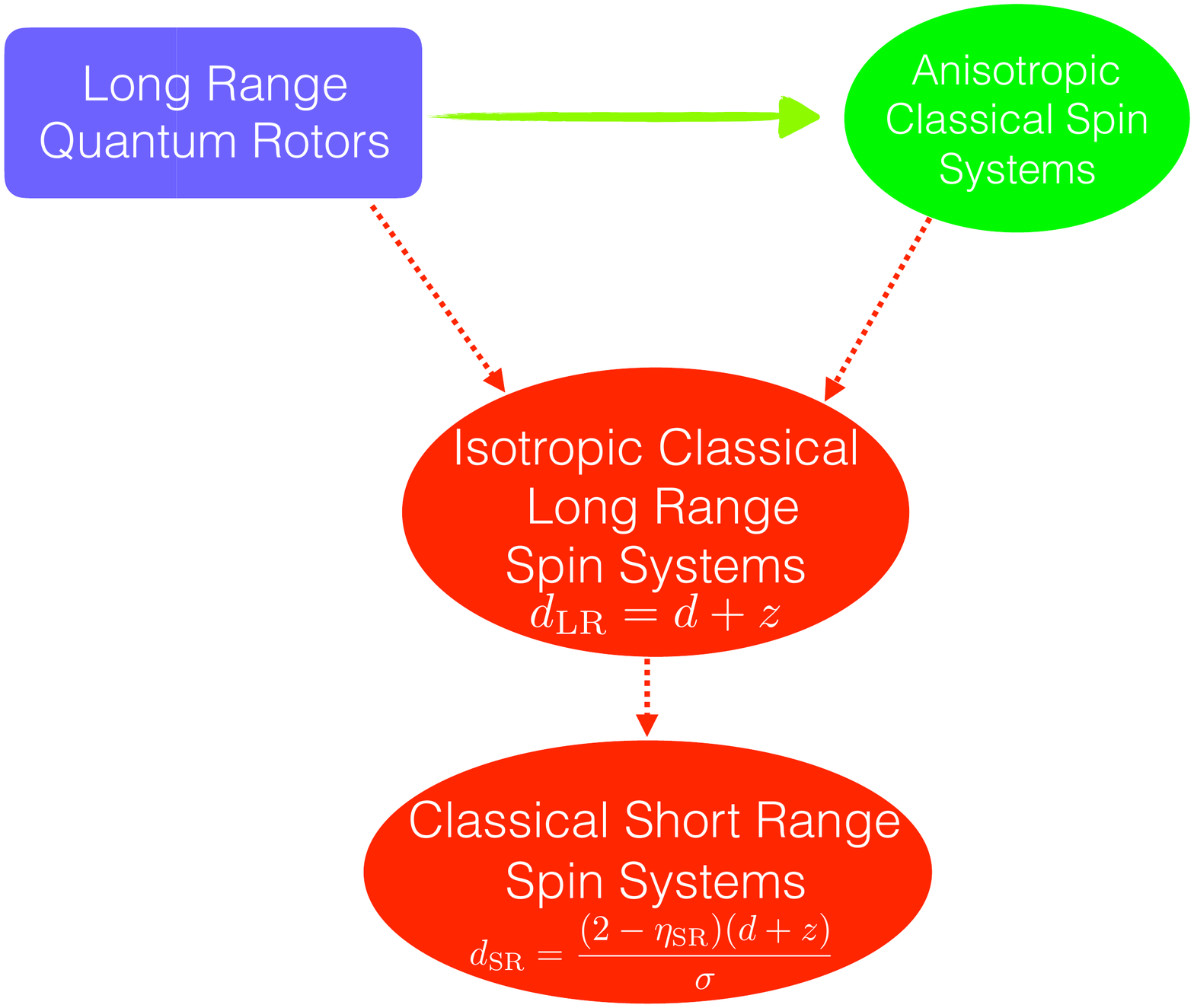}
\caption{Schematic representation of the relations between quantum and 
classical LR systems investigated in the paper. The green line indicates the exact mapping obtained via Trotter decomposition, while the red dotted lines indicate effective dimension relations, which, although not exact, produce very accurate results in the case of quantum LR models.}
\label{Fig4}
\end{figure*}

Such correspondance naturally emerges in our framework, even if
it appears not to be exact due to the difference 
in the kinetics sectors of the two models. This scenario is 
schematically drawn in figure \ref{Fig4}, where the dashed lines stand 
for a non-exact mapping. Indeed the coefficients proportional
to $\eta_{\tau}$ appearing
in the second and third lines of equation \eqref{Eq25} 
are substantially different from the ones of the classical case 
\cite{Defenu2015}.
%

These coefficients are regulator dependent
quantities whose precise value can be shown to not affect much the final result
for the critical exponents, indeed both coefficients are proportional to $\eta_{\tau}$ which
is never larger than $0.25$ and they can be shown to produce corrections to the critical
exponents which are of order $O(\eta_{\tau}^{2})$. We also explicitly solved the equations both in presence and absence of the two coefficients in the classical case \cite{Defenu2015} showing that the error arising from neglecting these two quantities is well below 1\% in all the $\sigma$ range, in the quantum case such error is expected to be  smaller. 

The results depicted in figure \ref{Fig3} 
are thus consistent with the naive expectation of equal universality
class for quantum systems in dimension $d$ and classical systems in dimension
$d+z$. As discussed before even at our simple approximation level such equivalence is spoiled
by the presence of the regulator dependent coefficients multiplying the non linear
term in \eqref{Eq25}. However, our results for the correlation length exponent can be exactly mapped into the ones of classical SR models via effective dimension relation \eqref{Eq31}, since we disregarded the regulator dependents factors on the second line of equation \eqref{Eq25}. These results are in good agreement with the MC data taken from \cite{Sperstad2012}, see figures \ref{Fig2} and \ref{Fig3}, showing,  once again, the reliability of the effective dimension approach \cite{Defenu2015}. 

The phase diagram of a LR system can be then 
represented in the $(d,\sigma)$ plane, as in figure \ref{Fig1}. For small 
enough values of the decay exponent $\sigma$ 
the system undergoes spontaneous symmetry breaking with mean field 
exponents
given by relations \eqref{eta_mf}, \eqref{z_mf} and \eqref{nu_mf}, cyan shaded region in figure \ref{Fig1}. For intermediate values
of $\sigma$ the system has peculiar LR exponents which continuously merge with the SR values at the $\sigma_{*}$ threshold.

In analogy with the classical case \cite{Angelini2014,Defenu2015},  
such threshold value
is given by $\sigma^{*}=2-\eta_{\rm{SR}}$, where $\eta_{\rm{SR}}$ 
is the anomalous dimension
found in the pure SR system.

It is worth noting that for continuous symmetries $N\geq 2$ it is 
also possible to identify a region, gray shaded area in figure \ref{Fig1b},  
where no phase transition is possible. Such identification does not 
have a counterpart in the discrete symmetry case of 
the Ising $N=1$, where no exact lower critical dimension is 
known even in the traditional SR case.

This theoretical picture advocates for more detailed comparison with 
ongoing experiments on LR power decaying interactions in quantum systems. 
Their understanding can pave the way to the comprehension of the role of 
dimension in quantum phase transitions and in the reliability of effective 
dimensional mappings.

Finally, we mention that a topic we did not address in detail 
in this paper is the emergence of the properties of the 
BKT phase transition in the $d=1$, $N=2$ limit. We think this is a deserving 
subject of investigation, also in connection with the properties 
of the $1$-dimensional XXZ with LR couplings 
\cite{Laflorencie2005,Bermudez2016}. 

\textit{Acknowledgements.} 
N. Defenu would like to thank I. Frerot for the useful discussion about the
relation between classical and quantum systems during the BEC
2016 conference. We are also pleased to acknowledge discussions 
with M. Ib\'a\~{n}ez Berganza, G. Parisi, L. Tagliacozzo and S. Whitlock.

\bibliography{Quantum_LR}

\begin{thebibliography}{57}%
\makeatletter
\providecommand \@ifxundefined [1]{%
 \@ifx{#1\undefined}
}%
\providecommand \@ifnum [1]{%
 \ifnum #1\expandafter \@firstoftwo
 \else \expandafter \@secondoftwo
 \fi
}%
\providecommand \@ifx [1]{%
 \ifx #1\expandafter \@firstoftwo
 \else \expandafter \@secondoftwo
 \fi
}%
\providecommand \natexlab [1]{#1}%
\providecommand \enquote  [1]{``#1''}%
\providecommand \bibnamefont  [1]{#1}%
\providecommand \bibfnamefont [1]{#1}%
\providecommand \citenamefont [1]{#1}%
\providecommand \href@noop [0]{\@secondoftwo}%
\providecommand \href [0]{\begingroup \@sanitize@url \@href}%
\providecommand \@href[1]{\@@startlink{#1}\@@href}%
\providecommand \@@href[1]{\endgroup#1\@@endlink}%
\providecommand \@sanitize@url [0]{\catcode `\\12\catcode `\$12\catcode
  `\&12\catcode `\#12\catcode `\^12\catcode `\_12\catcode `\%12\relax}%
\providecommand \@@startlink[1]{}%
\providecommand \@@endlink[0]{}%
\providecommand \url  [0]{\begingroup\@sanitize@url \@url }%
\providecommand \@url [1]{\endgroup\@href {#1}{\urlprefix }}%
\providecommand \urlprefix  [0]{URL }%
\providecommand \Eprint [0]{\href }%
\providecommand \doibase [0]{http://dx.doi.org/}%
\providecommand \selectlanguage [0]{\@gobble}%
\providecommand \bibinfo  [0]{\@secondoftwo}%
\providecommand \bibfield  [0]{\@secondoftwo}%
\providecommand \translation [1]{[#1]}%
\providecommand \BibitemOpen [0]{}%
\providecommand \bibitemStop [0]{}%
\providecommand \bibitemNoStop [0]{.\EOS\space}%
\providecommand \EOS [0]{\spacefactor3000\relax}%
\providecommand \BibitemShut  [1]{\csname bibitem#1\endcsname}%
\let\auto@bib@innerbib\@empty
\bibitem [{\citenamefont {Saffman}\ \emph {et~al.}(2009)\citenamefont
  {Saffman}, \citenamefont {Walker},\ and\ \citenamefont
  {Molmer}}]{Saffman2010}%
  \BibitemOpen
  \bibfield  {author} {\bibinfo {author} {\bibfnamefont {M.}~\bibnamefont
  {Saffman}}, \bibinfo {author} {\bibfnamefont {T.~G.}\ \bibnamefont {Walker}},
  \ and\ \bibinfo {author} {\bibfnamefont {K.}~\bibnamefont {Molmer}},\
  }\bibfield  {title} {\enquote {\bibinfo {title} {{Quantum information with
  Rydberg atoms}},}\ }\href {\doibase 10.1103/RevModPhys.82.2313} {\bibfield
  {journal} {\bibinfo  {journal} {Rev. Mod. Phys.}\ }\textbf {\bibinfo {volume}
  {82}},\ \bibinfo {pages} {2313--2363} (\bibinfo {year} {2009})},\ \Eprint
  {http://arxiv.org/abs/0909.4777} {arXiv:0909.4777} \BibitemShut {NoStop}%
\bibitem [{\citenamefont {Lahaye}\ \emph {et~al.}(2009)\citenamefont {Lahaye},
  \citenamefont {Menotti}, \citenamefont {Santos}, \citenamefont {Lewenstein},\
  and\ \citenamefont {Pfau}}]{Lahaye2009}%
  \BibitemOpen
  \bibfield  {author} {\bibinfo {author} {\bibfnamefont {T}~\bibnamefont
  {Lahaye}}, \bibinfo {author} {\bibfnamefont {C}~\bibnamefont {Menotti}},
  \bibinfo {author} {\bibfnamefont {L}~\bibnamefont {Santos}}, \bibinfo
  {author} {\bibfnamefont {M}~\bibnamefont {Lewenstein}}, \ and\ \bibinfo
  {author} {\bibfnamefont {T}~\bibnamefont {Pfau}},\ }\bibfield  {title}
  {\enquote {\bibinfo {title} {{The physics of dipolar bosonic quantum
  gases}},}\ }\href {\doibase 10.1088/0034-4885/72/12/126401} {\bibfield
  {journal} {\bibinfo  {journal} {Reports Prog. Phys.}\ }\textbf {\bibinfo
  {volume} {71}},\ \bibinfo {pages} {126401} (\bibinfo {year} {2009})},\
  \Eprint {http://arxiv.org/abs/0905.0386} {arXiv:0905.0386} \BibitemShut
  {NoStop}%
\bibitem [{\citenamefont {Ritsch}\ \emph {et~al.}(2013)\citenamefont {Ritsch},
  \citenamefont {Domokos}, \citenamefont {Brennecke},\ and\ \citenamefont
  {Esslinger}}]{Ritsch2013}%
  \BibitemOpen
  \bibfield  {author} {\bibinfo {author} {\bibfnamefont {Helmut}\ \bibnamefont
  {Ritsch}}, \bibinfo {author} {\bibfnamefont {Peter}\ \bibnamefont {Domokos}},
  \bibinfo {author} {\bibfnamefont {Ferdinand}\ \bibnamefont {Brennecke}}, \
  and\ \bibinfo {author} {\bibfnamefont {Tilman}\ \bibnamefont {Esslinger}},\
  }\bibfield  {title} {\enquote {\bibinfo {title} {{Cold atoms in
  cavity-generated dynamical optical potentials}},}\ }\href {\doibase
  10.1103/RevModPhys.85.553} {\bibfield  {journal} {\bibinfo  {journal} {Rev.
  Mod. Phys.}\ }\textbf {\bibinfo {volume} {85}},\ \bibinfo {pages} {553--601}
  (\bibinfo {year} {2013})}\BibitemShut {NoStop}%
\bibitem [{\citenamefont {Carr}\ \emph {et~al.}(2009)\citenamefont {Carr},
  \citenamefont {DeMille}, \citenamefont {Krems},\ and\ \citenamefont
  {Ye}}]{Carr2009}%
  \BibitemOpen
  \bibfield  {author} {\bibinfo {author} {\bibfnamefont {Lincoln~D}\
  \bibnamefont {Carr}}, \bibinfo {author} {\bibfnamefont {David}\ \bibnamefont
  {DeMille}}, \bibinfo {author} {\bibfnamefont {Roman~V}\ \bibnamefont
  {Krems}}, \ and\ \bibinfo {author} {\bibfnamefont {Jun}\ \bibnamefont {Ye}},\
  }\bibfield  {title} {\enquote {\bibinfo {title} {{Cold and ultracold
  molecules: science, technology and applications}},}\ }\href {\doibase
  10.1088/1367-2630/11/5/055049} {\bibfield  {journal} {\bibinfo  {journal}
  {New J. Phys.}\ }\textbf {\bibinfo {volume} {11}},\ \bibinfo {pages} {055049}
  (\bibinfo {year} {2009})}\BibitemShut {NoStop}%
\bibitem [{\citenamefont {Blatt}\ and\ \citenamefont {Roos}(2012)}]{Blatt2012}%
  \BibitemOpen
  \bibfield  {author} {\bibinfo {author} {\bibfnamefont {R.}~\bibnamefont
  {Blatt}}\ and\ \bibinfo {author} {\bibfnamefont {C.~F.}\ \bibnamefont
  {Roos}},\ }\bibfield  {title} {\enquote {\bibinfo {title} {{Quantum
  simulations with trapped ions}},}\ }\href {\doibase 10.1038/nphys2252}
  {\bibfield  {journal} {\bibinfo  {journal} {Nat. Phys.}\ }\textbf {\bibinfo
  {volume} {8}},\ \bibinfo {pages} {277--284} (\bibinfo {year}
  {2012})}\BibitemShut {NoStop}%
\bibitem [{\citenamefont {Bloch}\ \emph {et~al.}(2008)\citenamefont {Bloch},
  \citenamefont {Dalibard},\ and\ \citenamefont {Zwerger}}]{Bloch2008}%
  \BibitemOpen
  \bibfield  {author} {\bibinfo {author} {\bibfnamefont {Immanuel}\
  \bibnamefont {Bloch}}, \bibinfo {author} {\bibfnamefont {Jean}\ \bibnamefont
  {Dalibard}}, \ and\ \bibinfo {author} {\bibfnamefont {Wilhelm}\ \bibnamefont
  {Zwerger}},\ }\bibfield  {title} {\enquote {\bibinfo {title} {{Many-body
  physics with ultracold gases}},}\ }\href {\doibase 10.1103/RevModPhys.80.885}
  {\bibfield  {journal} {\bibinfo  {journal} {Rev. Mod. Phys.}\ }\textbf
  {\bibinfo {volume} {80}},\ \bibinfo {pages} {885--964} (\bibinfo {year}
  {2008})}\BibitemShut {NoStop}%
\bibitem [{\citenamefont {Britton}\ \emph {et~al.}(2012)\citenamefont
  {Britton}, \citenamefont {Sawyer}, \citenamefont {Keith}, \citenamefont
  {Wang}, \citenamefont {Freericks}, \citenamefont {Uys}, \citenamefont
  {Biercuk},\ and\ \citenamefont {Bollinger}}]{Britton2012}%
  \BibitemOpen
  \bibfield  {author} {\bibinfo {author} {\bibfnamefont {Joseph~W.}\
  \bibnamefont {Britton}}, \bibinfo {author} {\bibfnamefont {Brian~C.}\
  \bibnamefont {Sawyer}}, \bibinfo {author} {\bibfnamefont {Adam~C.}\
  \bibnamefont {Keith}}, \bibinfo {author} {\bibfnamefont {C.-C.~Joseph}\
  \bibnamefont {Wang}}, \bibinfo {author} {\bibfnamefont {James~K.}\
  \bibnamefont {Freericks}}, \bibinfo {author} {\bibfnamefont {Hermann}\
  \bibnamefont {Uys}}, \bibinfo {author} {\bibfnamefont {Michael~J.}\
  \bibnamefont {Biercuk}}, \ and\ \bibinfo {author} {\bibfnamefont {John~J.}\
  \bibnamefont {Bollinger}},\ }\bibfield  {title} {\enquote {\bibinfo {title}
  {{Engineered two-dimensional Ising interactions in a trapped-ion quantum
  simulator with hundreds of spins}},}\ }\href {\doibase 10.1038/nature10981}
  {\bibfield  {journal} {\bibinfo  {journal} {Nature}\ }\textbf {\bibinfo
  {volume} {484}},\ \bibinfo {pages} {489--492} (\bibinfo {year}
  {2012})}\BibitemShut {NoStop}%
\bibitem [{\citenamefont {Schau{\ss}}\ \emph {et~al.}(2012)\citenamefont
  {Schau{\ss}}, \citenamefont {Cheneau}, \citenamefont {Endres}, \citenamefont
  {Fukuhara}, \citenamefont {Hild}, \citenamefont {Omran}, \citenamefont
  {Pohl}, \citenamefont {Gross}, \citenamefont {Kuhr},\ and\ \citenamefont
  {Bloch}}]{Schauss2012}%
  \BibitemOpen
  \bibfield  {author} {\bibinfo {author} {\bibfnamefont {Peter}\ \bibnamefont
  {Schau{\ss}}}, \bibinfo {author} {\bibfnamefont {Marc}\ \bibnamefont
  {Cheneau}}, \bibinfo {author} {\bibfnamefont {Manuel}\ \bibnamefont
  {Endres}}, \bibinfo {author} {\bibfnamefont {Takeshi}\ \bibnamefont
  {Fukuhara}}, \bibinfo {author} {\bibfnamefont {Sebastian}\ \bibnamefont
  {Hild}}, \bibinfo {author} {\bibfnamefont {Ahmed}\ \bibnamefont {Omran}},
  \bibinfo {author} {\bibfnamefont {Thomas}\ \bibnamefont {Pohl}}, \bibinfo
  {author} {\bibfnamefont {Christian}\ \bibnamefont {Gross}}, \bibinfo {author}
  {\bibfnamefont {Stefan}\ \bibnamefont {Kuhr}}, \ and\ \bibinfo {author}
  {\bibfnamefont {Immanuel}\ \bibnamefont {Bloch}},\ }\bibfield  {title}
  {\enquote {\bibinfo {title} {{Observation of spatially ordered structures in
  a two-dimensional Rydberg gas}},}\ }\href {\doibase 10.1038/nature11596}
  {\bibfield  {journal} {\bibinfo  {journal} {Nature}\ }\textbf {\bibinfo
  {volume} {491}},\ \bibinfo {pages} {87--91} (\bibinfo {year}
  {2012})}\BibitemShut {NoStop}%
\bibitem [{\citenamefont {Aikawa}\ \emph {et~al.}(2012)\citenamefont {Aikawa},
  \citenamefont {Frisch}, \citenamefont {Mark}, \citenamefont {Baier},
  \citenamefont {Rietzler}, \citenamefont {Grimm},\ and\ \citenamefont
  {Ferlaino}}]{Aikawa2012}%
  \BibitemOpen
  \bibfield  {author} {\bibinfo {author} {\bibfnamefont {K.}~\bibnamefont
  {Aikawa}}, \bibinfo {author} {\bibfnamefont {A.}~\bibnamefont {Frisch}},
  \bibinfo {author} {\bibfnamefont {M.}~\bibnamefont {Mark}}, \bibinfo {author}
  {\bibfnamefont {S.}~\bibnamefont {Baier}}, \bibinfo {author} {\bibfnamefont
  {A.}~\bibnamefont {Rietzler}}, \bibinfo {author} {\bibfnamefont
  {R.}~\bibnamefont {Grimm}}, \ and\ \bibinfo {author} {\bibfnamefont
  {F.}~\bibnamefont {Ferlaino}},\ }\bibfield  {title} {\enquote {\bibinfo
  {title} {{Bose-Einstein Condensation of Erbium}},}\ }\href {\doibase
  10.1103/PhysRevLett.108.210401} {\bibfield  {journal} {\bibinfo  {journal}
  {Phys. Rev. Lett.}\ }\textbf {\bibinfo {volume} {108}},\ \bibinfo {pages}
  {210401} (\bibinfo {year} {2012})}\BibitemShut {NoStop}%
\bibitem [{\citenamefont {Lu}\ \emph {et~al.}(2012)\citenamefont {Lu},
  \citenamefont {Burdick},\ and\ \citenamefont {Lev}}]{Lu2012}%
  \BibitemOpen
  \bibfield  {author} {\bibinfo {author} {\bibfnamefont {Mingwu}\ \bibnamefont
  {Lu}}, \bibinfo {author} {\bibfnamefont {Nathaniel~Q.}\ \bibnamefont
  {Burdick}}, \ and\ \bibinfo {author} {\bibfnamefont {Benjamin~L.}\
  \bibnamefont {Lev}},\ }\bibfield  {title} {\enquote {\bibinfo {title}
  {{Quantum Degenerate Dipolar Fermi Gas}},}\ }\href {\doibase
  10.1103/PhysRevLett.108.215301} {\bibfield  {journal} {\bibinfo  {journal}
  {Phys. Rev. Lett.}\ }\textbf {\bibinfo {volume} {108}},\ \bibinfo {pages}
  {215301} (\bibinfo {year} {2012})}\BibitemShut {NoStop}%
\bibitem [{\citenamefont {Yan}\ \emph {et~al.}(2013)\citenamefont {Yan},
  \citenamefont {Moses}, \citenamefont {Gadway}, \citenamefont {Covey},
  \citenamefont {Hazzard}, \citenamefont {Rey}, \citenamefont {Jin},\ and\
  \citenamefont {Ye}}]{Yan2013}%
  \BibitemOpen
  \bibfield  {author} {\bibinfo {author} {\bibfnamefont {Bo}~\bibnamefont
  {Yan}}, \bibinfo {author} {\bibfnamefont {Steven~A.}\ \bibnamefont {Moses}},
  \bibinfo {author} {\bibfnamefont {Bryce}\ \bibnamefont {Gadway}}, \bibinfo
  {author} {\bibfnamefont {Jacob~P.}\ \bibnamefont {Covey}}, \bibinfo {author}
  {\bibfnamefont {Kaden R.~A.}\ \bibnamefont {Hazzard}}, \bibinfo {author}
  {\bibfnamefont {Ana~Maria}\ \bibnamefont {Rey}}, \bibinfo {author}
  {\bibfnamefont {Deborah~S.}\ \bibnamefont {Jin}}, \ and\ \bibinfo {author}
  {\bibfnamefont {Jun}\ \bibnamefont {Ye}},\ }\bibfield  {title} {\enquote
  {\bibinfo {title} {{Observation of dipolar spin-exchange interactions with
  lattice-confined polar molecules}},}\ }\href {\doibase 10.1038/nature12483}
  {\bibfield  {journal} {\bibinfo  {journal} {Nature}\ }\textbf {\bibinfo
  {volume} {501}},\ \bibinfo {pages} {521--525} (\bibinfo {year}
  {2013})}\BibitemShut {NoStop}%
\bibitem [{\citenamefont {Islam}\ \emph {et~al.}(2013)\citenamefont {Islam},
  \citenamefont {Senko}, \citenamefont {Campbell}, \citenamefont {Korenblit},
  \citenamefont {Smith}, \citenamefont {Lee}, \citenamefont {Edwards},
  \citenamefont {Wang}, \citenamefont {Freericks},\ and\ \citenamefont
  {Monroe}}]{Islam2013}%
  \BibitemOpen
  \bibfield  {author} {\bibinfo {author} {\bibfnamefont {R.}~\bibnamefont
  {Islam}}, \bibinfo {author} {\bibfnamefont {C.}~\bibnamefont {Senko}},
  \bibinfo {author} {\bibfnamefont {W.~C.}\ \bibnamefont {Campbell}}, \bibinfo
  {author} {\bibfnamefont {S.}~\bibnamefont {Korenblit}}, \bibinfo {author}
  {\bibfnamefont {J.}~\bibnamefont {Smith}}, \bibinfo {author} {\bibfnamefont
  {A.}~\bibnamefont {Lee}}, \bibinfo {author} {\bibfnamefont {E.~E.}\
  \bibnamefont {Edwards}}, \bibinfo {author} {\bibfnamefont {C.-C. J. C.~J.}\
  \bibnamefont {Wang}}, \bibinfo {author} {\bibfnamefont {J.~K.}\ \bibnamefont
  {Freericks}}, \ and\ \bibinfo {author} {\bibfnamefont {C.}~\bibnamefont
  {Monroe}},\ }\bibfield  {title} {\enquote {\bibinfo {title} {{Emergence and
  Frustration of Magnetism with Variable-Range Interactions in a Quantum
  Simulator}},}\ }\href {\doibase 10.1126/science.1232296} {\bibfield
  {journal} {\bibinfo  {journal} {Science (80-. ).}\ }\textbf {\bibinfo
  {volume} {340}},\ \bibinfo {pages} {583--587} (\bibinfo {year}
  {2013})}\BibitemShut {NoStop}%
\bibitem [{\citenamefont {Richerme}\ \emph {et~al.}(2014)\citenamefont
  {Richerme}, \citenamefont {Gong}, \citenamefont {Lee}, \citenamefont {Senko},
  \citenamefont {Smith}, \citenamefont {Foss-Feig}, \citenamefont {Michalakis},
  \citenamefont {Gorshkov},\ and\ \citenamefont {Monroe}}]{Richerme2014}%
  \BibitemOpen
  \bibfield  {author} {\bibinfo {author} {\bibfnamefont {Philip}\ \bibnamefont
  {Richerme}}, \bibinfo {author} {\bibfnamefont {Zhe-Xuan}\ \bibnamefont
  {Gong}}, \bibinfo {author} {\bibfnamefont {Aaron}\ \bibnamefont {Lee}},
  \bibinfo {author} {\bibfnamefont {Crystal}\ \bibnamefont {Senko}}, \bibinfo
  {author} {\bibfnamefont {Jacob}\ \bibnamefont {Smith}}, \bibinfo {author}
  {\bibfnamefont {Michael}\ \bibnamefont {Foss-Feig}}, \bibinfo {author}
  {\bibfnamefont {Spyridon}\ \bibnamefont {Michalakis}}, \bibinfo {author}
  {\bibfnamefont {Alexey~V.}\ \bibnamefont {Gorshkov}}, \ and\ \bibinfo
  {author} {\bibfnamefont {Christopher}\ \bibnamefont {Monroe}},\ }\bibfield
  {title} {\enquote {\bibinfo {title} {{Non-local propagation of correlations
  in quantum systems with long-range interactions}},}\ }\href {\doibase
  10.1038/nature13450} {\bibfield  {journal} {\bibinfo  {journal} {Nature}\
  }\textbf {\bibinfo {volume} {511}},\ \bibinfo {pages} {198--201} (\bibinfo
  {year} {2014})},\ \Eprint {http://arxiv.org/abs/1401.5088} {arXiv:1401.5088}
  \BibitemShut {NoStop}%
\bibitem [{\citenamefont {Jurcevic}\ \emph {et~al.}(2014)\citenamefont
  {Jurcevic}, \citenamefont {Lanyon}, \citenamefont {Hauke}, \citenamefont
  {Hempel}, \citenamefont {Zoller}, \citenamefont {Blatt},\ and\ \citenamefont
  {Roos}}]{Jurcevic2014}%
  \BibitemOpen
  \bibfield  {author} {\bibinfo {author} {\bibfnamefont {P.}~\bibnamefont
  {Jurcevic}}, \bibinfo {author} {\bibfnamefont {B.~P.}\ \bibnamefont
  {Lanyon}}, \bibinfo {author} {\bibfnamefont {P.}~\bibnamefont {Hauke}},
  \bibinfo {author} {\bibfnamefont {C.}~\bibnamefont {Hempel}}, \bibinfo
  {author} {\bibfnamefont {P.}~\bibnamefont {Zoller}}, \bibinfo {author}
  {\bibfnamefont {R.}~\bibnamefont {Blatt}}, \ and\ \bibinfo {author}
  {\bibfnamefont {C.~F.}\ \bibnamefont {Roos}},\ }\bibfield  {title} {\enquote
  {\bibinfo {title} {{Quasiparticle engineering and entanglement propagation in
  a quantum many-body system}},}\ }\href {\doibase 10.1038/nature13461}
  {\bibfield  {journal} {\bibinfo  {journal} {Nature}\ }\textbf {\bibinfo
  {volume} {511}},\ \bibinfo {pages} {202--205} (\bibinfo {year}
  {2014})}\BibitemShut {NoStop}%
\bibitem [{\citenamefont {Douglas}\ \emph {et~al.}(2015)\citenamefont
  {Douglas}, \citenamefont {Habibian}, \citenamefont {Hung}, \citenamefont
  {Gorshkov}, \citenamefont {Kimble},\ and\ \citenamefont
  {Chang}}]{Douglas2015}%
  \BibitemOpen
  \bibfield  {author} {\bibinfo {author} {\bibfnamefont {J.~S.}\ \bibnamefont
  {Douglas}}, \bibinfo {author} {\bibfnamefont {H.}~\bibnamefont {Habibian}},
  \bibinfo {author} {\bibfnamefont {C.-L.}\ \bibnamefont {Hung}}, \bibinfo
  {author} {\bibfnamefont {A.~V.}\ \bibnamefont {Gorshkov}}, \bibinfo {author}
  {\bibfnamefont {H.~J.}\ \bibnamefont {Kimble}}, \ and\ \bibinfo {author}
  {\bibfnamefont {D.~E.}\ \bibnamefont {Chang}},\ }\bibfield  {title} {\enquote
  {\bibinfo {title} {{Quantum many-body models with cold atoms coupled to
  photonic crystals}},}\ }\href {\doibase 10.1038/nphoton.2015.57} {\bibfield
  {journal} {\bibinfo  {journal} {Nat. Photonics}\ }\textbf {\bibinfo {volume}
  {9}},\ \bibinfo {pages} {326--331} (\bibinfo {year} {2015})}\BibitemShut
  {NoStop}%
\bibitem [{\citenamefont {Schempp}\ \emph {et~al.}(2015)\citenamefont
  {Schempp}, \citenamefont {G{\"{u}}nter}, \citenamefont {W{\"{u}}ster},
  \citenamefont {Weidem{\"{u}}ller},\ and\ \citenamefont
  {Whitlock}}]{Schempp2015}%
  \BibitemOpen
  \bibfield  {author} {\bibinfo {author} {\bibfnamefont {H.}~\bibnamefont
  {Schempp}}, \bibinfo {author} {\bibfnamefont {G.}~\bibnamefont
  {G{\"{u}}nter}}, \bibinfo {author} {\bibfnamefont {S.}~\bibnamefont
  {W{\"{u}}ster}}, \bibinfo {author} {\bibfnamefont {M.}~\bibnamefont
  {Weidem{\"{u}}ller}}, \ and\ \bibinfo {author} {\bibfnamefont
  {S.}~\bibnamefont {Whitlock}},\ }\bibfield  {title} {\enquote {\bibinfo
  {title} {{Correlated Exciton Transport in Rydberg-Dressed-Atom Spin
  Chains}},}\ }\href {\doibase 10.1103/PhysRevLett.115.093002} {\bibfield
  {journal} {\bibinfo  {journal} {Phys. Rev. Lett.}\ }\textbf {\bibinfo
  {volume} {115}},\ \bibinfo {pages} {093002} (\bibinfo {year} {2015})},\
  \Eprint {http://arxiv.org/abs/1504.01892} {arXiv:1504.01892} \BibitemShut
  {NoStop}%
\bibitem [{\citenamefont {Landig}\ \emph {et~al.}(2015)\citenamefont {Landig},
  \citenamefont {Brennecke}, \citenamefont {Mottl}, \citenamefont {Donner},\
  and\ \citenamefont {Esslinger}}]{Landig2015}%
  \BibitemOpen
  \bibfield  {author} {\bibinfo {author} {\bibfnamefont {Renate}\ \bibnamefont
  {Landig}}, \bibinfo {author} {\bibfnamefont {Ferdinand}\ \bibnamefont
  {Brennecke}}, \bibinfo {author} {\bibfnamefont {Rafael}\ \bibnamefont
  {Mottl}}, \bibinfo {author} {\bibfnamefont {Tobias}\ \bibnamefont {Donner}},
  \ and\ \bibinfo {author} {\bibfnamefont {Tilman}\ \bibnamefont {Esslinger}},\
  }\bibfield  {title} {\enquote {\bibinfo {title} {{Measuring the dynamic
  structure factor of a quantum gas undergoing a structural phase
  transition}},}\ }\href {\doibase 10.1038/ncomms8046} {\bibfield  {journal}
  {\bibinfo  {journal} {Nat. Commun.}\ }\textbf {\bibinfo {volume} {6}},\
  \bibinfo {pages} {7046} (\bibinfo {year} {2015})},\ \Eprint
  {http://arxiv.org/abs/1503.05565} {arXiv:1503.05565} \BibitemShut {NoStop}%
\bibitem [{\citenamefont {Landig}\ \emph {et~al.}(2016)\citenamefont {Landig},
  \citenamefont {Hruby}, \citenamefont {Dogra}, \citenamefont {Landini},
  \citenamefont {Mottl}, \citenamefont {Donner},\ and\ \citenamefont
  {Esslinger}}]{Landig2015a}%
  \BibitemOpen
  \bibfield  {author} {\bibinfo {author} {\bibfnamefont {Renate}\ \bibnamefont
  {Landig}}, \bibinfo {author} {\bibfnamefont {Lorenz}\ \bibnamefont {Hruby}},
  \bibinfo {author} {\bibfnamefont {Nishant}\ \bibnamefont {Dogra}}, \bibinfo
  {author} {\bibfnamefont {Manuele}\ \bibnamefont {Landini}}, \bibinfo {author}
  {\bibfnamefont {Rafael}\ \bibnamefont {Mottl}}, \bibinfo {author}
  {\bibfnamefont {Tobias}\ \bibnamefont {Donner}}, \ and\ \bibinfo {author}
  {\bibfnamefont {Tilman}\ \bibnamefont {Esslinger}},\ }\bibfield  {title}
  {\enquote {\bibinfo {title} {{Quantum phases from competing short- and
  long-range interactions in an optical lattice}},}\ }\href {\doibase
  10.1038/nature17409} {\bibfield  {journal} {\bibinfo  {journal} {Nature}\
  }\textbf {\bibinfo {volume} {532}},\ \bibinfo {pages} {476--479} (\bibinfo
  {year} {2016})},\ \Eprint {http://arxiv.org/abs/1511.00007}
  {arXiv:1511.00007} \BibitemShut {NoStop}%
\bibitem [{\citenamefont {Martinez}\ \emph {et~al.}(2016)\citenamefont
  {Martinez}, \citenamefont {Muschik}, \citenamefont {Schindler}, \citenamefont
  {Nigg}, \citenamefont {Erhard}, \citenamefont {Heyl}, \citenamefont {Hauke},
  \citenamefont {Dalmonte}, \citenamefont {Monz}, \citenamefont {Zoller},\ and\
  \citenamefont {Blatt}}]{Martinez2016}%
  \BibitemOpen
  \bibfield  {author} {\bibinfo {author} {\bibfnamefont {Esteban~A.}\
  \bibnamefont {Martinez}}, \bibinfo {author} {\bibfnamefont {Christine~A.}\
  \bibnamefont {Muschik}}, \bibinfo {author} {\bibfnamefont {Philipp}\
  \bibnamefont {Schindler}}, \bibinfo {author} {\bibfnamefont {Daniel}\
  \bibnamefont {Nigg}}, \bibinfo {author} {\bibfnamefont {Alexander}\
  \bibnamefont {Erhard}}, \bibinfo {author} {\bibfnamefont {Markus}\
  \bibnamefont {Heyl}}, \bibinfo {author} {\bibfnamefont {Philipp}\
  \bibnamefont {Hauke}}, \bibinfo {author} {\bibfnamefont {Marcello}\
  \bibnamefont {Dalmonte}}, \bibinfo {author} {\bibfnamefont {Thomas}\
  \bibnamefont {Monz}}, \bibinfo {author} {\bibfnamefont {Peter}\ \bibnamefont
  {Zoller}}, \ and\ \bibinfo {author} {\bibfnamefont {Rainer}\ \bibnamefont
  {Blatt}},\ }\bibfield  {title} {\enquote {\bibinfo {title} {{Real-time
  dynamics of lattice gauge theories with a few-qubit quantum computer}},}\
  }\href {\doibase 10.1038/nature18318} {\bibfield  {journal} {\bibinfo
  {journal} {Nature}\ }\textbf {\bibinfo {volume} {534}},\ \bibinfo {pages}
  {516--519} (\bibinfo {year} {2016})}\BibitemShut {NoStop}%
\bibitem [{\citenamefont {Codello}\ \emph
  {et~al.}(2015{\natexlab{a}})\citenamefont {Codello}, \citenamefont {Defenu},\
  and\ \citenamefont {D'Odorico}}]{Dauxois2010}%
  \BibitemOpen
  \bibfield  {author} {\bibinfo {author} {\bibfnamefont {Alessandro}\
  \bibnamefont {Codello}}, \bibinfo {author} {\bibfnamefont {Nicol{\'{o}}}\
  \bibnamefont {Defenu}}, \ and\ \bibinfo {author} {\bibfnamefont {Giulio}\
  \bibnamefont {D'Odorico}},\ }\bibfield  {title} {\enquote {\bibinfo {title}
  {{Critical exponents of O (N) models in fractional dimensions}},}\ }\href
  {\doibase 10.1103/PhysRevD.91.105003} {\bibfield  {journal} {\bibinfo
  {journal} {Phys. Rev. D - Part. Fields, Gravit. Cosmol.}\ }\textbf {\bibinfo
  {volume} {91}},\ \bibinfo {pages} {105003} (\bibinfo {year}
  {2015}{\natexlab{a}})},\ \Eprint {http://arxiv.org/abs/1410.3308v2}
  {arXiv:1410.3308v2} \BibitemShut {NoStop}%
\bibitem [{\citenamefont {Campa}\ \emph {et~al.}(2014)\citenamefont {Campa},
  \citenamefont {Dauxois}, \citenamefont {Fanelli},\ and\ \citenamefont
  {Ruffo}}]{Campa2014}%
  \BibitemOpen
  \bibfield  {author} {\bibinfo {author} {\bibfnamefont {A}~\bibnamefont
  {Campa}}, \bibinfo {author} {\bibfnamefont {Thierry}\ \bibnamefont
  {Dauxois}}, \bibinfo {author} {\bibfnamefont {D}~\bibnamefont {Fanelli}}, \
  and\ \bibinfo {author} {\bibfnamefont {S}~\bibnamefont {Ruffo}},\ }\href
  {https://books.google.it/books?hl=it{\&}lr={\&}id=-vg{\_}BAAAQBAJ{\&}oi=fnd{\&}pg=PP1{\&}dq=campa+2014{\&}ots=5iGTE9gpFJ{\&}sig=uNdrmNB9p4LutUBfNzNgvWLDRRc
  https://books.google.com/books?id=AIw{\_}BAAAQBAJ{\&}pgis=1} {\emph {\bibinfo
  {title} {{Physics of Long-Range Interacting Systems}}}}\ (\bibinfo {year}
  {2014})\ p.\ \bibinfo {pages} {410}\BibitemShut {NoStop}%
\bibitem [{\citenamefont {Sachdev}(2011)}]{Sachdev2011}%
  \BibitemOpen
  \bibfield  {author} {\bibinfo {author} {\bibfnamefont {Subir}\ \bibnamefont
  {Sachdev}},\ }\href {\doibase 10.1017/CBO9780511973765} {\emph {\bibinfo
  {title} {{Quantum Phase Transitions}}}}\ (\bibinfo  {publisher} {Cambridge
  University Press},\ \bibinfo {address} {Cambridge},\ \bibinfo {year} {2011})\
  p.\ \bibinfo {pages} {517},\ \Eprint {http://arxiv.org/abs/9811058}
  {arXiv:9811058 [cond-mat]} \BibitemShut {NoStop}%
\bibitem [{\citenamefont {Defenu}\ \emph {et~al.}(2015)\citenamefont {Defenu},
  \citenamefont {Trombettoni},\ and\ \citenamefont {Codello}}]{Defenu2015}%
  \BibitemOpen
  \bibfield  {author} {\bibinfo {author} {\bibfnamefont {Nicol{\'{o}}}\
  \bibnamefont {Defenu}}, \bibinfo {author} {\bibfnamefont {Andrea}\
  \bibnamefont {Trombettoni}}, \ and\ \bibinfo {author} {\bibfnamefont
  {Alessandro}\ \bibnamefont {Codello}},\ }\bibfield  {title} {\enquote
  {\bibinfo {title} {{Fixed-point structure and effective fractional
  dimensionality for O( N ) models with long-range interactions}},}\ }\href
  {\doibase 10.1103/PhysRevE.92.052113} {\bibfield  {journal} {\bibinfo
  {journal} {Phys. Rev. E}\ }\textbf {\bibinfo {volume} {92}},\ \bibinfo
  {pages} {052113} (\bibinfo {year} {2015})},\ \Eprint
  {http://arxiv.org/abs/1211.3991} {arXiv:1211.3991} \BibitemShut {NoStop}%
\bibitem [{\citenamefont {Defenu}\ \emph {et~al.}(2016)\citenamefont {Defenu},
  \citenamefont {Trombettoni},\ and\ \citenamefont {Ruffo}}]{Defenu2016}%
  \BibitemOpen
  \bibfield  {author} {\bibinfo {author} {\bibfnamefont {Nicol{\`{o}}}\
  \bibnamefont {Defenu}}, \bibinfo {author} {\bibfnamefont {Andrea}\
  \bibnamefont {Trombettoni}}, \ and\ \bibinfo {author} {\bibfnamefont
  {Stefano}\ \bibnamefont {Ruffo}},\ }\bibfield  {title} {\enquote {\bibinfo
  {title} {{Anisotropic Long-Range Spin Systems}},}\ }\href {\doibase
  10.1103/PhysRevB.94.224411} {\bibfield  {journal} {\bibinfo  {journal} {Phys.
  Rev. B}\ }\textbf {\bibinfo {volume} {94}},\ \bibinfo {pages} {224411}
  (\bibinfo {year} {2016})},\ \Eprint {http://arxiv.org/abs/1606.07756}
  {arXiv:1606.07756} \BibitemShut {NoStop}%
\bibitem [{\citenamefont {Thouless}(1969)}]{Thouless1969}%
  \BibitemOpen
  \bibfield  {author} {\bibinfo {author} {\bibfnamefont {D.~J.}\ \bibnamefont
  {Thouless}},\ }\bibfield  {title} {\enquote {\bibinfo {title} {{Long-Range
  Order in One-Dimensional Ising Systems}},}\ }\href {\doibase
  10.1103/PhysRev.187.732} {\bibfield  {journal} {\bibinfo  {journal} {Phys.
  Rev.}\ }\textbf {\bibinfo {volume} {187}},\ \bibinfo {pages} {732--733}
  (\bibinfo {year} {1969})}\BibitemShut {NoStop}%
\bibitem [{\citenamefont {Yuval}\ and\ \citenamefont
  {Anderson}(1970)}]{Yuval1970}%
  \BibitemOpen
  \bibfield  {author} {\bibinfo {author} {\bibfnamefont {G.}~\bibnamefont
  {Yuval}}\ and\ \bibinfo {author} {\bibfnamefont {P.~W.}\ \bibnamefont
  {Anderson}},\ }\bibfield  {title} {\enquote {\bibinfo {title} {{Exact results
  for the kondo problem: One-body theory and extension to finite
  temperature}},}\ }\href {\doibase 10.1103/PhysRevB.1.1522} {\bibfield
  {journal} {\bibinfo  {journal} {Phys. Rev. B}\ }\textbf {\bibinfo {volume}
  {1}},\ \bibinfo {pages} {1522--1528} (\bibinfo {year} {1970})}\BibitemShut
  {NoStop}%
\bibitem [{\citenamefont {Cardy}(1981)}]{Cardy1981}%
  \BibitemOpen
  \bibfield  {author} {\bibinfo {author} {\bibfnamefont {J~L}\ \bibnamefont
  {Cardy}},\ }\bibfield  {title} {\enquote {\bibinfo {title} {{One-dimensional
  models with $1/r^{2}$
  interactions}},}\ }\href {\doibase 10.1088/0305-4470/14/6/017} {\bibfield
  {journal} {\bibinfo  {journal} {J. Phys. A. Math. Gen.}\ }\textbf {\bibinfo
  {volume} {14}},\ \bibinfo {pages} {1407--1415} (\bibinfo {year}
  {1981})}\BibitemShut {NoStop}%
\bibitem [{\citenamefont {Dutta}\ and\ \citenamefont
  {Bhattacharjee}(2001)}]{Dutta2001}%
  \BibitemOpen
  \bibfield  {author} {\bibinfo {author} {\bibfnamefont {Amit}\ \bibnamefont
  {Dutta}}\ and\ \bibinfo {author} {\bibfnamefont {J.}~\bibnamefont
  {Bhattacharjee}},\ }\bibfield  {title} {\enquote {\bibinfo {title} {{Phase
  transitions in the quantum Ising and rotor models with a long-range
  interaction}},}\ }\href {\doibase 10.1103/PhysRevB.64.184106} {\bibfield
  {journal} {\bibinfo  {journal} {Phys. Rev. B}\ }\textbf {\bibinfo {volume}
  {64}},\ \bibinfo {pages} {1--7} (\bibinfo {year} {2001})}\BibitemShut
  {NoStop}%
\bibitem [{\citenamefont {Sak}(1973)}]{Sak1973}%
  \BibitemOpen
  \bibfield  {author} {\bibinfo {author} {\bibfnamefont {J.}~\bibnamefont
  {Sak}},\ }\bibfield  {title} {\enquote {\bibinfo {title} {{Recursion
  relations and fixed points for ferromagnets with long-range interactions}},}\
  }\href {\doibase 10.1103/PhysRevB.8.281} {\bibfield  {journal} {\bibinfo
  {journal} {Phys. Rev. B}\ }\textbf {\bibinfo {volume} {8}},\ \bibinfo {pages}
  {281--285} (\bibinfo {year} {1973})}\BibitemShut {NoStop}%
\bibitem [{\citenamefont {Picco}(2012)}]{Picco2012}%
  \BibitemOpen
  \bibfield  {author} {\bibinfo {author} {\bibfnamefont {Marco}\ \bibnamefont
  {Picco}},\ }\bibfield  {title} {\enquote {\bibinfo {title} {{Critical
  behavior of the Ising model with long range interactions}},}\ }\href
  {http://arxiv.org/abs/1207.1018} {\  (\bibinfo {year} {2012})},\ \Eprint
  {http://arxiv.org/abs/1207.1018} {arXiv:1207.1018} \BibitemShut {NoStop}%
\bibitem [{\citenamefont {Blanchard}\ \emph {et~al.}(2013)\citenamefont
  {Blanchard}, \citenamefont {Picco},\ and\ \citenamefont
  {Rajabpour}}]{Blanchard2013}%
  \BibitemOpen
  \bibfield  {author} {\bibinfo {author} {\bibfnamefont {Thibault}\
  \bibnamefont {Blanchard}}, \bibinfo {author} {\bibfnamefont {Marco}\
  \bibnamefont {Picco}}, \ and\ \bibinfo {author} {\bibfnamefont {M.~A.}\
  \bibnamefont {Rajabpour}},\ }\bibfield  {title} {\enquote {\bibinfo {title}
  {{Influence of long-range interactions on the critical behavior of the Ising
  model}},}\ }\href {\doibase 10.1209/0295-5075/101/56003} {\bibfield
  {journal} {\bibinfo  {journal} {EPL}\ }\textbf {\bibinfo {volume} {101}},\
  \bibinfo {pages} {56003} (\bibinfo {year} {2013})},\ \Eprint
  {http://arxiv.org/abs/1211.6758} {arXiv:1211.6758} \BibitemShut {NoStop}%
\bibitem [{\citenamefont {Brezin}\ \emph {et~al.}(2014)\citenamefont {Brezin},
  \citenamefont {Parisi},\ and\ \citenamefont {Ricci-Tersenghi}}]{Brezin2014}%
  \BibitemOpen
  \bibfield  {author} {\bibinfo {author} {\bibfnamefont {E.}~\bibnamefont
  {Brezin}}, \bibinfo {author} {\bibfnamefont {G.}~\bibnamefont {Parisi}}, \
  and\ \bibinfo {author} {\bibfnamefont {F.}~\bibnamefont {Ricci-Tersenghi}},\
  }\bibfield  {title} {\enquote {\bibinfo {title} {{The Crossover Region
  Between Long-Range and Short-Range Interactions for the Critical
  Exponents}},}\ }\href {\doibase 10.1007/s10955-014-1081-0} {\bibfield
  {journal} {\bibinfo  {journal} {J. Stat. Phys.}\ }\textbf {\bibinfo {volume}
  {157}},\ \bibinfo {pages} {855--868} (\bibinfo {year} {2014})}\BibitemShut
  {NoStop}%
\bibitem [{\citenamefont {Luijten}\ and\ \citenamefont
  {Bl{\"{o}}te}(2002)}]{Luijten2002}%
  \BibitemOpen
  \bibfield  {author} {\bibinfo {author} {\bibfnamefont {Erik}\ \bibnamefont
  {Luijten}}\ and\ \bibinfo {author} {\bibfnamefont {Henk W.~J.}\ \bibnamefont
  {Bl{\"{o}}te}},\ }\bibfield  {title} {\enquote {\bibinfo {title} {{Boundary
  between Long-Range and Short-Range Critical Behavior in Systems with
  Algebraic Interactions}},}\ }\href {\doibase 10.1103/PhysRevLett.89.025703}
  {\bibfield  {journal} {\bibinfo  {journal} {Phys. Rev. Lett.}\ }\textbf
  {\bibinfo {volume} {89}},\ \bibinfo {pages} {25703} (\bibinfo {year}
  {2002})}\BibitemShut {NoStop}%
\bibitem [{\citenamefont {Angelini}\ \emph {et~al.}(2014)\citenamefont
  {Angelini}, \citenamefont {Parisi},\ and\ \citenamefont
  {Ricci-Tersenghi}}]{Angelini2014}%
  \BibitemOpen
  \bibfield  {author} {\bibinfo {author} {\bibfnamefont {Maria~Chiara}\
  \bibnamefont {Angelini}}, \bibinfo {author} {\bibfnamefont {Giorgio}\
  \bibnamefont {Parisi}}, \ and\ \bibinfo {author} {\bibfnamefont {Federico}\
  \bibnamefont {Ricci-Tersenghi}},\ }\bibfield  {title} {\enquote {\bibinfo
  {title} {{Relations between short-range and long-range Ising models}},}\
  }\href {\doibase 10.1103/PhysRevE.89.062120} {\bibfield  {journal} {\bibinfo
  {journal} {Phys. Rev. E}\ }\textbf {\bibinfo {volume} {89}},\ \bibinfo
  {pages} {062120} (\bibinfo {year} {2014})}\BibitemShut {NoStop}%
\bibitem [{\citenamefont {Mori}(2010)}]{Mori2010}%
  \BibitemOpen
  \bibfield  {author} {\bibinfo {author} {\bibfnamefont {Takashi}\ \bibnamefont
  {Mori}},\ }\bibfield  {title} {\enquote {\bibinfo {title} {{Analysis of the
  exactness of mean-field theory in long-range interacting systems}},}\ }\href
  {\doibase 10.1103/PhysRevE.82.060103} {\bibfield  {journal} {\bibinfo
  {journal} {Phys. Rev. E}\ }\textbf {\bibinfo {volume} {82}},\ \bibinfo
  {pages} {060103} (\bibinfo {year} {2010})},\ \Eprint
  {http://arxiv.org/abs/1004.3622} {arXiv:1004.3622} \BibitemShut {NoStop}%
\bibitem [{\citenamefont {Horita}\ \emph {et~al.}(2016)\citenamefont {Horita},
  \citenamefont {Suwa},\ and\ \citenamefont {Todo}}]{Horita2016}%
  \BibitemOpen
  \bibfield  {author} {\bibinfo {author} {\bibfnamefont {Toshiki}\ \bibnamefont
  {Horita}}, \bibinfo {author} {\bibfnamefont {Hidemaro}\ \bibnamefont {Suwa}},
  \ and\ \bibinfo {author} {\bibfnamefont {Synge}\ \bibnamefont {Todo}},\
  }\bibfield  {title} {\enquote {\bibinfo {title} {{Upper and Lower Critical
  Decay Exponents of Ising Ferromagnets with Long-range Interaction}},}\ }\href
  {\doibase 10.1103/PhysRevE.95.012143} {\bibfield  {journal} {\bibinfo
  {journal} {Phys. Rev. E}\ }\textbf {\bibinfo {volume} {95}},\ \bibinfo
  {pages} {012143} (\bibinfo {year} {2016})},\ \Eprint
  {http://arxiv.org/abs/1605.09496} {arXiv:1605.09496} \BibitemShut {NoStop}%
\bibitem [{\citenamefont {Gori}\ \emph {et~al.}(2017)\citenamefont {Gori},
  \citenamefont {Michelangeli}, \citenamefont {Defenu},\ and\ \citenamefont
  {Trombettoni}}]{Gori2016}%
  \BibitemOpen
  \bibfield  {author} {\bibinfo {author} {\bibfnamefont {G.}~\bibnamefont
  {Gori}}, \bibinfo {author} {\bibfnamefont {M.}~\bibnamefont {Michelangeli}},
  \bibinfo {author} {\bibfnamefont {N.}~\bibnamefont {Defenu}}, \ and\ \bibinfo
  {author} {\bibfnamefont {A.}~\bibnamefont {Trombettoni}},\ }\bibfield
  {title} {\enquote {\bibinfo {title} {{One-dimensional long-range percolation:
  a numerical study}},}\ }\href {\doibase 10.1103/PhysRevE.96.012108}
  {\bibfield  {journal} {\bibinfo  {journal} {Phys. Rev. E}\ }\textbf {\bibinfo
  {volume} {96}},\ \bibinfo {pages} {012108} (\bibinfo {year} {2017})},\
  \Eprint {http://arxiv.org/abs/1610.00200} {arXiv:1610.00200} \BibitemShut
  {NoStop}%
\bibitem [{\citenamefont {Codello}\ and\ \citenamefont
  {D'Odorico}(2013)}]{Codello2013}%
  \BibitemOpen
  \bibfield  {author} {\bibinfo {author} {\bibfnamefont {Alessandro}\
  \bibnamefont {Codello}}\ and\ \bibinfo {author} {\bibfnamefont {Giulio}\
  \bibnamefont {D'Odorico}},\ }\bibfield  {title} {\enquote {\bibinfo {title}
  {{O(N) -Universality Classes and the Mermin-Wagner Theorem}},}\ }\href
  {\doibase 10.1103/PhysRevLett.110.141601} {\bibfield  {journal} {\bibinfo
  {journal} {Phys. Rev. Lett.}\ }\textbf {\bibinfo {volume} {110}},\ \bibinfo
  {pages} {141601} (\bibinfo {year} {2013})}\BibitemShut {NoStop}%
\bibitem [{\citenamefont {Codello}\ \emph
  {et~al.}(2015{\natexlab{b}})\citenamefont {Codello}, \citenamefont {Defenu},
  \citenamefont {D'Odorico},\ and\ \citenamefont {D'Odorico}}]{Codello2015}%
  \BibitemOpen
  \bibfield  {author} {\bibinfo {author} {\bibfnamefont {Alessandro}\
  \bibnamefont {Codello}}, \bibinfo {author} {\bibfnamefont {Nicol{\'{o}}}\
  \bibnamefont {Defenu}}, \bibinfo {author} {\bibfnamefont {Giulio}\
  \bibnamefont {D'Odorico}}, \ and\ \bibinfo {author} {\bibfnamefont {Giulio}\
  \bibnamefont {D'Odorico}},\ }\bibfield  {title} {\enquote {\bibinfo {title}
  {{Critical exponents of $O(N)$ models in fractional dimensions}},}\ }\href
  {\doibase 10.1103/PhysRevD.91.105003} {\bibfield  {journal} {\bibinfo
  {journal} {Phys. Rev. D}\ }\textbf {\bibinfo {volume} {91}},\ \bibinfo
  {pages} {105003} (\bibinfo {year} {2015}{\natexlab{b}})}\BibitemShut
  {NoStop}%
\bibitem [{\citenamefont {Uhlenbeck}\ \emph {et~al.}(1963)\citenamefont
  {Uhlenbeck}, \citenamefont {Hemmer},\ and\ \citenamefont {Kac}}]{Kac1963}%
  \BibitemOpen
  \bibfield  {author} {\bibinfo {author} {\bibfnamefont {G.~E.}\ \bibnamefont
  {Uhlenbeck}}, \bibinfo {author} {\bibfnamefont {P.~C.}\ \bibnamefont
  {Hemmer}}, \ and\ \bibinfo {author} {\bibfnamefont {M.}~\bibnamefont {Kac}},\
  }\bibfield  {title} {\enquote {\bibinfo {title} {{On the van der Waals Theory
  of the Vapor-Liquid Equilibrium. I. Discussion of a One-Dimensional
  Model}},}\ }\href {\doibase 10.1063/1.1703946} {\bibfield  {journal}
  {\bibinfo  {journal} {J. Math. Phys.}\ }\textbf {\bibinfo {volume} {4}},\
  \bibinfo {pages} {216} (\bibinfo {year} {1963})}\BibitemShut {NoStop}%
\bibitem [{\citenamefont {Behan}\ \emph
  {et~al.}(2017{\natexlab{a}})\citenamefont {Behan}, \citenamefont {Rastelli},
  \citenamefont {Rychkov},\ and\ \citenamefont {Zan}}]{Behan2017}%
  \BibitemOpen
  \bibfield  {author} {\bibinfo {author} {\bibfnamefont {Connor}\ \bibnamefont
  {Behan}}, \bibinfo {author} {\bibfnamefont {Leonardo}\ \bibnamefont
  {Rastelli}}, \bibinfo {author} {\bibfnamefont {Slava}\ \bibnamefont
  {Rychkov}}, \ and\ \bibinfo {author} {\bibfnamefont {Bernardo}\ \bibnamefont
  {Zan}},\ }\bibfield  {title} {\enquote {\bibinfo {title} {{A scaling theory
  for the long-range to short-range crossover and an infrared duality}},}\
  }\href {http://arxiv.org/abs/1703.03430 https://arxiv.org/pdf/1703.03430.pdf
  http://arxiv.org/abs/1703.05325} {\bibfield  {journal} {\bibinfo  {journal}
  {To Appear}\ } (\bibinfo {year} {2017}{\natexlab{a}})},\ \Eprint
  {http://arxiv.org/abs/1703.05325v1} {arXiv:1703.05325v1} \BibitemShut
  {NoStop}%
\bibitem [{\citenamefont {Behan}\ \emph
  {et~al.}(2017{\natexlab{b}})\citenamefont {Behan}, \citenamefont {Rastelli},
  \citenamefont {Rychkov},\ and\ \citenamefont {Zan}}]{Behan2017a}%
  \BibitemOpen
  \bibfield  {author} {\bibinfo {author} {\bibfnamefont {Connor}\ \bibnamefont
  {Behan}}, \bibinfo {author} {\bibfnamefont {Leonardo}\ \bibnamefont
  {Rastelli}}, \bibinfo {author} {\bibfnamefont {Slava}\ \bibnamefont
  {Rychkov}}, \ and\ \bibinfo {author} {\bibfnamefont {Bernardo}\ \bibnamefont
  {Zan}},\ }\bibfield  {title} {\enquote {\bibinfo {title} {{Long-range
  critical exponents near the short-range crossover}},}\ }\href {\doibase
  10.1103/PhysRevLett.118.241601} {\bibfield  {journal} {\bibinfo  {journal}
  {eprint arXiv:1703.03430}\ } (\bibinfo {year} {2017}{\natexlab{b}}),\
  10.1103/PhysRevLett.118.241601},\ \Eprint {http://arxiv.org/abs/1703.03430}
  {arXiv:1703.03430} \BibitemShut {NoStop}%
\bibitem [{\citenamefont {Tetradis}\ and\ \citenamefont
  {Wetterich}(1992)}]{Tetradis1992}%
  \BibitemOpen
  \bibfield  {author} {\bibinfo {author} {\bibfnamefont {N.}~\bibnamefont
  {Tetradis}}\ and\ \bibinfo {author} {\bibfnamefont {C.}~\bibnamefont
  {Wetterich}},\ }\bibfield  {title} {\enquote {\bibinfo {title} {{Scale
  dependence of the average potential around the maximum in $\phi$4
  theories}},}\ }\href {\doibase 10.1016/0550-3213(92)90676-3} {\bibfield
  {journal} {\bibinfo  {journal} {Nucl. Phys. B}\ }\textbf {\bibinfo {volume}
  {383}},\ \bibinfo {pages} {197--217} (\bibinfo {year} {1992})}\BibitemShut
  {NoStop}%
\bibitem [{\citenamefont {Morris}(1994)}]{Morris1994}%
  \BibitemOpen
  \bibfield  {author} {\bibinfo {author} {\bibfnamefont {Tim~R.}\ \bibnamefont
  {Morris}},\ }\bibfield  {title} {\enquote {\bibinfo {title} {{Derivative
  expansion of the exact renormalization group}},}\ }\href {\doibase
  10.1016/0370-2693(94)90767-6} {\bibfield  {journal} {\bibinfo  {journal}
  {Phys. Lett. B}\ }\textbf {\bibinfo {volume} {329}},\ \bibinfo {pages}
  {241--248} (\bibinfo {year} {1994})},\ \Eprint {http://arxiv.org/abs/0108031}
  {arXiv:0108031 [hep-th]} \BibitemShut {NoStop}%
\bibitem [{\citenamefont {Berges}\ \emph {et~al.}(2002)\citenamefont {Berges},
  \citenamefont {Tetradis},\ and\ \citenamefont {Wetterich}}]{Berges2002}%
  \BibitemOpen
  \bibfield  {author} {\bibinfo {author} {\bibfnamefont {J{\"{u}}rgen}\
  \bibnamefont {Berges}}, \bibinfo {author} {\bibfnamefont {Nikolaos}\
  \bibnamefont {Tetradis}}, \ and\ \bibinfo {author} {\bibfnamefont {Christof}\
  \bibnamefont {Wetterich}},\ }\bibfield  {title} {\enquote {\bibinfo {title}
  {{Non-perturbative renormalization flow in quantum field theory and
  statistical physics}},}\ }\href {\doibase 10.1016/S0370-1573(01)00098-9}
  {\bibfield  {journal} {\bibinfo  {journal} {Phys. Rep.}\ }\textbf {\bibinfo
  {volume} {363}},\ \bibinfo {pages} {223--386} (\bibinfo {year}
  {2002})}\BibitemShut {NoStop}%
\bibitem [{\citenamefont {Mussardo}(2010)}]{Mussardo2010}%
  \BibitemOpen
  \bibfield  {author} {\bibinfo {author} {\bibfnamefont {G.}~\bibnamefont
  {Mussardo}},\ }\href@noop {} {\emph {\bibinfo {title} {{Statistical field
  theory : an introduction to exactly solved models in statistical physics}}}}\
  (\bibinfo  {publisher} {Oxford University Press},\ \bibinfo {year} {2010})\
  p.\ \bibinfo {pages} {755}\BibitemShut {NoStop}%
\bibitem [{\citenamefont {Wetterich}(1993)}]{Wetterich1993}%
  \BibitemOpen
  \bibfield  {author} {\bibinfo {author} {\bibfnamefont {Christof}\
  \bibnamefont {Wetterich}},\ }\bibfield  {title} {\enquote {\bibinfo {title}
  {{Exact evolution equation for the effective potential}},}\ }\href {\doibase
  10.1016/0370-2693(93)90726-X} {\bibfield  {journal} {\bibinfo  {journal}
  {Phys. Lett. B}\ }\textbf {\bibinfo {volume} {301}},\ \bibinfo {pages}
  {90--94} (\bibinfo {year} {1993})}\BibitemShut {NoStop}%
\bibitem [{\citenamefont {Joyce}(1966)}]{Joyce1966}%
  \BibitemOpen
  \bibfield  {author} {\bibinfo {author} {\bibfnamefont {G.~S.}\ \bibnamefont
  {Joyce}},\ }\bibfield  {title} {\enquote {\bibinfo {title} {{Spherical model
  with long-range ferromagnetic interactions}},}\ }\href {\doibase
  10.1103/PhysRev.146.349} {\bibfield  {journal} {\bibinfo  {journal} {Phys.
  Rev.}\ }\textbf {\bibinfo {volume} {146}},\ \bibinfo {pages} {349--358}
  (\bibinfo {year} {1966})}\BibitemShut {NoStop}%
\bibitem [{\citenamefont {Dutta}(2003)}]{Dutta2003}%
  \BibitemOpen
  \bibfield  {author} {\bibinfo {author} {\bibfnamefont {Amit}\ \bibnamefont
  {Dutta}},\ }\bibfield  {title} {\enquote {\bibinfo {title} {{Effect of
  long-range interactions on the pure and random quantum Ising transitions}},}\
  }in\ \href {\doibase 10.1016/S0378-4371(02)01404-8} {\emph {\bibinfo
  {booktitle} {Phys. A Stat. Mech. its Appl.}}},\ Vol.\ \bibinfo {volume}
  {318}\ (\bibinfo {year} {2003})\ pp.\ \bibinfo {pages} {63--71}\BibitemShut
  {NoStop}%
\bibitem [{\citenamefont {Sperstad}\ \emph {et~al.}(2012)\citenamefont
  {Sperstad}, \citenamefont {Stiansen},\ and\ \citenamefont
  {Sudb{\o}}}]{Sperstad2012}%
  \BibitemOpen
  \bibfield  {author} {\bibinfo {author} {\bibfnamefont {Iver~Bakken}\
  \bibnamefont {Sperstad}}, \bibinfo {author} {\bibfnamefont {Einar~B.}\
  \bibnamefont {Stiansen}}, \ and\ \bibinfo {author} {\bibfnamefont {Asle}\
  \bibnamefont {Sudb{\o}}},\ }\bibfield  {title} {\enquote {\bibinfo {title}
  {{Quantum criticality in spin chains with non-Ohmic dissipation}},}\ }\href
  {\doibase 10.1103/PhysRevB.85.214302} {\bibfield  {journal} {\bibinfo
  {journal} {Phys. Rev. B}\ }\textbf {\bibinfo {volume} {85}},\ \bibinfo
  {pages} {214302} (\bibinfo {year} {2012})},\ \Eprint
  {http://arxiv.org/abs/1204.2538} {arXiv:1204.2538} \BibitemShut {NoStop}%
\bibitem [{\citenamefont {Br??zin}\ and\ \citenamefont
  {Zinn-Justin}(1976)}]{Brezin1976}%
  \BibitemOpen
  \bibfield  {author} {\bibinfo {author} {\bibfnamefont {E.}~\bibnamefont
  {Br\'ezin}}\ and\ \bibinfo {author} {\bibfnamefont {J.}~\bibnamefont
  {Zinn-Justin}},\ }\bibfield  {title} {\enquote {\bibinfo {title}
  {{Renormalization of the nonlinear $\sigma$ model in 2+$\varepsilon$ dimensions-application to
  the heisenberg ferromagnets}},}\ }\href {\doibase 10.1103/PhysRevLett.36.691}
  {\bibfield  {journal} {\bibinfo  {journal} {Phys. Rev. Lett.}\ }\textbf
  {\bibinfo {volume} {36}},\ \bibinfo {pages} {691--694} (\bibinfo {year}
  {1976})}\BibitemShut {NoStop}%
\bibitem [{\citenamefont {Fey}\ and\ \citenamefont {Schmidt}(2016)}]{Fey2016}%
  \BibitemOpen
  \bibfield  {author} {\bibinfo {author} {\bibfnamefont {Sebastian}\
  \bibnamefont {Fey}}\ and\ \bibinfo {author} {\bibfnamefont {Kai~Phillip}\
  \bibnamefont {Schmidt}},\ }\bibfield  {title} {\enquote {\bibinfo {title}
  {{Critical behavior of quantum magnets with long-range interactions in the
  thermodynamic limit}},}\ }\href {\doibase 10.1103/PhysRevB.94.075156}
  {\bibfield  {journal} {\bibinfo  {journal} {Phys. Rev. B}\ }\textbf {\bibinfo
  {volume} {94}},\ \bibinfo {pages} {075156} (\bibinfo {year}
  {2016})}\BibitemShut {NoStop}%
\bibitem [{\citenamefont {Morris}\ and\ \citenamefont
  {Turner}(1998)}]{Morris1998a}%
  \BibitemOpen
  \bibfield  {author} {\bibinfo {author} {\bibfnamefont {Tim~R}\ \bibnamefont
  {Morris}}\ and\ \bibinfo {author} {\bibfnamefont {Michael~D}\ \bibnamefont
  {Turner}},\ }\bibfield  {title} {\enquote {\bibinfo {title} {{Derivative
  Expansion of the Renormalization Group in O(N) Scalar Field Theory}},}\
  }\href {\doibase 10.1016/S0550-3213(97)00640-8} {\bibfield  {journal}
  {\bibinfo  {journal} {Nucl. Phys. B}\ }\textbf {\bibinfo {volume} {509}},\
  \bibinfo {pages} {637--661} (\bibinfo {year} {1998})},\ \Eprint
  {http://arxiv.org/abs/9704202} {arXiv:9704202 [hep-th]} \BibitemShut
  {NoStop}%
\bibitem [{\citenamefont {Nieuwenhuizen}(1995)}]{Nieuwenhuizen1995}%
  \BibitemOpen
  \bibfield  {author} {\bibinfo {author} {\bibfnamefont {Th.~M.}\ \bibnamefont
  {Nieuwenhuizen}},\ }\bibfield  {title} {\enquote {\bibinfo {title} {{Quantum
  Description of Spherical Spins}},}\ }\href {\doibase
  10.1103/PhysRevLett.74.4293} {\bibfield  {journal} {\bibinfo  {journal}
  {Phys. Rev. Lett.}\ }\textbf {\bibinfo {volume} {74}},\ \bibinfo {pages}
  {4293--4296} (\bibinfo {year} {1995})}\BibitemShut {NoStop}%
\bibitem [{\citenamefont {Vojta}(1996)}]{Vojta1996}%
  \BibitemOpen
  \bibfield  {author} {\bibinfo {author} {\bibfnamefont {Thomas}\ \bibnamefont
  {Vojta}},\ }\bibfield  {title} {\enquote {\bibinfo {title} {{Quantum version
  of a spherical model: Crossover from quantum to classical critical
  behavior}},}\ }\href {\doibase 10.1103/PhysRevB.53.710} {\bibfield  {journal}
  {\bibinfo  {journal} {Phys. Rev. B}\ }\textbf {\bibinfo {volume} {53}},\
  \bibinfo {pages} {710} (\bibinfo {year} {1996})}\BibitemShut {NoStop}%
\bibitem [{\citenamefont {Laflorencie}\ \emph {et~al.}(2005)\citenamefont
  {Laflorencie}, \citenamefont {Affleck},\ and\ \citenamefont
  {Berciu}}]{Laflorencie2005}%
  \BibitemOpen
  \bibfield  {author} {\bibinfo {author} {\bibfnamefont {Nicolas}\ \bibnamefont
  {Laflorencie}}, \bibinfo {author} {\bibfnamefont {Ian}\ \bibnamefont
  {Affleck}}, \ and\ \bibinfo {author} {\bibfnamefont {Mona}\ \bibnamefont
  {Berciu}},\ }\bibfield  {title} {\enquote {\bibinfo {title} {{Critical
  phenomena and quantum phase transition in long range Heisenberg
  antiferromagnetic chains}},}\ }\href {\doibase
  10.1088/1742-5468/2005/12/P12001} {\bibfield  {journal} {\bibinfo  {journal}
  {J. Stat. Mech. Theory Exp.}\ }\textbf {\bibinfo {volume} {2005}},\ \bibinfo
  {pages} {P12001--P12001} (\bibinfo {year} {2005})},\ \Eprint
  {http://arxiv.org/abs/0509390} {arXiv:0509390 [cond-mat]} \BibitemShut
  {NoStop}%
\bibitem [{\citenamefont {Bermudez}\ \emph {et~al.}(2016)\citenamefont
  {Bermudez}, \citenamefont {Tagliacozzo}, \citenamefont {Sierra},\ and\
  \citenamefont {Richerme}}]{Bermudez2016}%
  \BibitemOpen
  \bibfield  {author} {\bibinfo {author} {\bibfnamefont {A.}~\bibnamefont
  {Bermudez}}, \bibinfo {author} {\bibfnamefont {L.}~\bibnamefont
  {Tagliacozzo}}, \bibinfo {author} {\bibfnamefont {G.}~\bibnamefont {Sierra}},
  \ and\ \bibinfo {author} {\bibfnamefont {P.}~\bibnamefont {Richerme}},\
  }\bibfield  {title} {\enquote {\bibinfo {title} {{Long-range Heisenberg
  models in quasi-periodically driven crystals of trapped ions}},}\ }\href
  {\doibase 10.1103/PhysRevB.95.024431} {\bibfield  {journal} {\bibinfo
  {journal} {Phys. Rev. B}\ }\textbf {\bibinfo {volume} {95}},\ \bibinfo
  {pages} {024431} (\bibinfo {year} {2016})},\ \Eprint
  {http://arxiv.org/abs/1607.03337} {arXiv:1607.03337} \BibitemShut {NoStop}%
\end{thebibliography}%




\end{document}